\documentclass[twocolumn,epjc3]{svjour3}  
\smartqed  % flush right qed marks, e.g. at end of proof
\RequirePackage{graphicx}
\RequirePackage[colorlinks,citecolor=blue,urlcolor=blue,linkcolor=blue]{hyperref}
\journalname{Eur. Phys. J. C}

\RequirePackage{lineno}\newdimen\linenumbersep\linenumbersep=2pt
\usepackage[T1]{fontenc}
\usepackage[utf8]{inputenc}
\usepackage{subfigure}
\usepackage{mathtools}
\usepackage{lineno}
\usepackage{array, booktabs, color, eurosym, longtable, pbox, pdfpages, siunitx, url, verbatim}

\usepackage{xspace}
\usepackage{wrapfig}
\usepackage{multirow}
\usepackage{blindtext}

\usepackage{babel}
\usepackage[font=small,labelfont=bf]{caption}

\newcommand{\doi}[1]{\href{http://dx.doi.org/#1}{\texttt{doi:#1}}}

\newcommand{\Ricochet}{\textsc{Ricochet}}

\makeatletter
\let\cl@chapter\undefined
\makeatletter
\usepackage[capitalise]{cleveref}

\usepackage{booktabs}
\usepackage{multirow}
\usepackage{textgreek}
\usepackage{siunitx}
\usepackage{xspace}
\usepackage{physics}

\makeatletter

\begin{document}

%%%%%%%%%%%%%%%%%%
%\linenumbers    %%   scott added this while reviewing, just highlighting here so it's obvious what to remove
%%%%%%%%%%%%%%%%%%

%\begin{frontmatter}

\title{Fast neutron background characterization of the future \Ricochet{} experiment at the ILL research nuclear reactor}% Force line breaks with \\

%
% need to make sure the affiliations get sorted / listed in a way
% that isn't too awkward
%
\author{
{C.~Augier}\thanksref{a}\and
{G.~Baulieu}\thanksref{a}\and
{V.~Belov}\thanksref{h}\and
{L.~Berge}\thanksref{b}\and
{J.~Billard}\thanksref{a,e2}\and
{G.~Bres}\thanksref{g}\and
{J-.L.~Bret}\thanksref{g}\and
{A.~Broniatowski}\thanksref{b}\and
{M.~Calvo}\thanksref{g}\and
{A.~Cazes}\thanksref{a}\and
{D.~Chaize}\thanksref{a}\and
{M.~Chapellier}\thanksref{b}\and
{L.~Chaplinsky}\thanksref{f}\and
{G.~Chemin}\thanksref{c}\and
{R.~Chen}\thanksref{d}\and
{J.~Colas}\thanksref{a}\and
{M.~De Jesus}\thanksref{a}\and
{P.~de Marcillac}\thanksref{b}\and
{L.~Dumoulin}\thanksref{b}\and
{O.~Exshaw}\thanksref{g}\and
{S.~Ferriol}\thanksref{a}\and
{E.~Figueroa-Feliciano}\thanksref{d}\and
{J.-B.~Filippini}\thanksref{a}\and
{J. A.~Formaggio}\thanksref{e}\and
{S.~Fuard}\thanksref{k}\and
{J.~Gascon}\thanksref{a}\and
{A.~Giuliani}\thanksref{b}\and
{J.~Goupy}\thanksref{g}\and
{C.~Goy}\thanksref{c}\and
{C.~Guerin}\thanksref{a}\and
{E.~Guy}\thanksref{a}\and
{P.~Harrington}\thanksref{e}\and
{S. T.~Heine}\thanksref{e}\and
{S. A.~Hertel}\thanksref{f}\and
{M.~Heusch}\thanksref{c}\and
{C. F.~Hirjibehedin }\thanksref{i}\and
{Z.~Hong}\thanksref{j}\and
{J.-C.~Ianigro}\thanksref{a}\and
{Y.~Jin}\thanksref{l}\and
{J. P.~Johnston}\thanksref{e}\and
{A.~Juillard}\thanksref{a}\and
{D.~Karaivanov}\thanksref{h}\and
{S.~Kazarcev}\thanksref{h}\and
{J.~Lamblin}\thanksref{c}\and
{H.~Lattaud}\thanksref{a}\and
{M.~Li}\thanksref{e}\and
{A.~Lubashevskiy}\thanksref{h}\and
{S.~Marnieros}\thanksref{b}\and
{D. W.~Mayer}\thanksref{e}\and
{J.~Minet}\thanksref{g}\and
{D.~Misiak}\thanksref{a}\and
{J-.L.~Mocellin}\thanksref{g}\and
{A.~Monfardini}\thanksref{g}\and
{F.~Mounier}\thanksref{a}\and
{W. D.~Oliver}\thanksref{i}\and
{E.~Olivieri}\thanksref{b}\and
{C.~Oriol}\thanksref{b}\and
{P. K.~Patel}\thanksref{f}\and
{E.~Perbet}\thanksref{c}\and
{H. D.~Pinckney}\thanksref{f}\and
{D.~Poda}\thanksref{b}\and
{D.~Ponomarev}\thanksref{h}\and
{F.~Rarbi}\thanksref{c}\and
{J.-S.~Real}\thanksref{c}\and
{T.~Redon}\thanksref{b}\and
{A.~Robert}\thanksref{k}\and
{S.~Rozov}\thanksref{h}\and
{I.~Rozova}\thanksref{h}\and
{T.~Salagnac}\thanksref{a}\and
{V.~Sanglard}\thanksref{a}\and
{B.~Schmidt}\thanksref{d}\and
{Ye.~Shevchik}\thanksref{h}\and
{V.~Sibille}\thanksref{a,e,e1}\and
{T.~Soldner}\thanksref{k}\and
{J.~Stachurska}\thanksref{e}\and
{A.~Stutz}\thanksref{c}\and
{L.~Vagneron}\thanksref{a}\and
{W.~Van De Pontseele}\thanksref{e}\and
{F.~Vezzu}\thanksref{c}\and
{S.~Weber}\thanksref{i}\and
{L.~Winslow}\thanksref{e}\and
{E.~Yakushev}\thanksref{h}\and
{D.~Zinatulina}\thanksref{h} 
 -- the \Ricochet{} Collaboration
}

\institute{
{Univ Lyon, Université Lyon 1, CNRS/IN2P3, IP2I-Lyon, F-69622, Villeurbanne, France}\label{a}\and
{Université Paris-Saclay, CNRS/IN2P3, IJCLab, 91405 Orsay, France}\label{b}\and
{Univ. Grenoble Alpes, CNRS, Grenoble INP, LPSC-IN2P3, Grenoble, France 38000}\label{c}\and
{Department of Physics and Astronomy, Northwestern University, IL, USA}\label{d}\and
{Laboratory for Nuclear Science, Massachusetts Institute of Technology, Cambridge, MA, USA 02139}\label{e}\and
{Department of Physics, University of Massachusetts at Amherst, Amherst, MA, USA 02139}\label{f}\and
{Univ. Grenoble Alpes, CNRS, Grenoble INP, Institut Néel, 38000 , Grenoble, France 38000}\label{g}\and
{Department of Nuclear Spectroscopy and Radiochemistry, Laboratory of Nuclear Problems, JINR, Dubna, Moscow Region, Russia 141980}\label{h}\and
{Lincoln Laboratory, Lexington, MA, USA}\label{i}\and
{Department of Physics, University of Toronto, ON, Canada M5S 1A7}\label{j}\and
{Institut Laue Langevin, Grenoble, France 38042}\label{k}\and
{C2N, CNRS, Univ. Paris-Saclay, Palaiseau, France 91120}\label{l}
}
%Email for the corresponding author

\thankstext{e1}{e-mail: vsibille@mit.edu}
\thankstext{e2}{e-mail: j.billard@ipnl.in2p3.fr}


%\keywords{Only keywords from JINST's keywords list please}

%\collaboration[c]{on behalf of the Ricochet Collaboration}

% the TOC can be omitted, of course, but especially as we're drafting and editing, it could be very helpful to see the overall scope
%\tableofcontents

\maketitle

\begin{abstract}
The  future \Ricochet{} experiment aims at searching for new physics in the electroweak sector by providing a high precision measurement of the Coherent Elastic Neutrino-Nucleus Scattering (CENNS) process down to the sub-100~eV nuclear recoil energy range. The experiment will deploy a kg-scale low-energy-threshold detector array combining Ge and Zn target crystals 8.8 meters away from the 58 MW research nuclear reactor core of the Institut Laue Langevin (ILL) in Grenoble, France. Currently, the \Ricochet{} collaboration is characterizing  the  backgrounds at its future experimental site in order to optimize the experiment's shielding design. The most threatening background component, which cannot be actively rejected by particle identification, consists of keV-scale neutron-induced nuclear recoils. These initial fast neutrons are generated by the reactor core and surrounding experiments (reactogenics), and by the cosmic rays producing primary neutrons and muon-induced neutrons in the surrounding materials. In this paper, we present the \Ricochet{} neutron background characterization using $^3$He proportional counters which exhibit a high sensitivity to thermal, epithermal and fast neutrons. We compare these measurements to the \Ricochet{} Geant4 simulations to validate our reactogenic and cosmogenic neutron background estimations. Eventually, we present our estimated neutron background for the future \Ricochet{} experiment and the resulting CENNS detection significance.
\end{abstract}

%\flushbottom

\section{Introduction}
\label{sec:ricochet-intro}

Coherent elastic neutrino-nucleus scattering (CENNS) was predicted in 1974~\cite{Freedman:1973yd} and observed experimentally for the first time in 2017~\cite{Akimov:2017ade}. This elastic scattering process, inducing nuclear recoils of a few keV at most, proceeds via the neutral weak current and benefits from a coherent enhancement proportional to the square of the number of neutrons~\cite{Freedman:1973yd}, suggesting that even a kg-scale experiment, located in the proximity of a research or commercial nuclear reactor, can observe a sizable neutrino signal. The search for physics beyond the Standard Model with CENNS requires to measure with the highest level of precision the sub-100~eV energy range of the induced nuclear recoils, as most new physics signatures induce energy spectral distortions in this energy region~\cite{Billard:2018jnl}. These include for instance the existence of sterile neutrinos and of new mediators that could be related to the long lasting Dark Matter problem, and the possibility of Non Standard Interactions that would dramatically affect our understanding of the electroweak sector. 

Thanks to its exceptionally rich science program, CENNS has led to significant worldwide experimental efforts over the last decades, with several ongoing and planned dedicated experiments based on a host of techniques. Most of these experiments are, or will be, located at nuclear reactor sites producing low-energy neutrinos with mean energies of about 3~MeV: CONNIE using Si-based CCDs~\cite{Aguilar-Arevalo:2016khx}; TEXONO~\cite{Kerman:2016jqp}, NuGEN~\cite{Belov:2015ufh}, and CONUS~\cite{Bonet:2020awv} using ionization-based Ge semiconductors; and MINER~\cite{Agnolet:2016zir}, NuCLEUS~\cite{Strauss:2017cuu}, and \Ricochet{}~\cite{Ricochet:2021rjo} using cryogenic detectors. Only the COHERENT experiment~\cite{Akimov:2017ade,Akimov:2020pdx} is looking at higher neutrino energies of about 30~MeV in average produced by the Spallation Neutron Source (SNS) in Oak Ridge, and
experiments are planned at the European Spallation Source (ESS) in Lund~\cite{Baxter:2019mcx}. 

The \Ricochet{} experiment seeks to utilize a kg-scale cryogenic detector payload combining Zn and Ge target crystals with sub-100~eV energy threshold and particle identification capabilities down to the energy threshold to reject the dominating gamma-induced electronic recoil background. Such identification will be achieved thanks to the double heat-and-ionization measurement with the semiconducting Ge target, and pulse shape discrimination in the superconducting  Zn crystals. In this context, the neutron-induced nuclear recoils are therefore expected to be the limiting background to the future \Ricochet{} experiment which will be located near the nuclear reactor of the Institut Laue Langevin (ILL). The close proximity to the reactor core comes at the cost of an additional reactor-correlated fast neutron background, called reactogenic neutrons, which could mimic a CENNS signal in the Ge and Zn target detectors hence limiting the expected \Ricochet{} CENNS sensitivity at ILL.

In this paper we present our fast neutron background characterization of the ILL-H7 site, where \Ricochet{} will be installed, and its implication on the expected background levels of the future \Ricochet{} experiment. To do so, we compare data taken with a $^3$He proportional counter sensitive to both thermal and fast neutrons with Geant4 simulations. Additionally, to further assess the robustness of the presented method we also characterized the cosmogenic neutron background at the {\it Institut de Physique des 2 Infinis de Lyon} (IP2I) cryogenic test facility, where Ge bolometers with particle identification capabilities have been operated~\cite{misiak:tel-03328713,billard:tel-03259707}. We show that this low-radioactivity $^3$He proportional counter is well-suited to constrain the fast neutron background at the future \Ricochet{} experiment. In light of these results, we conclude with the \Ricochet{} shielding optimization and the anticipated nuclear recoil background induced by reactogenic and cosmogenic neutrons.

\section{The Ricochet experiment}
\label{sec:Ricochet}
\begin{figure*}
\begin{center}
\includegraphics[width=0.44\textwidth,angle=0]{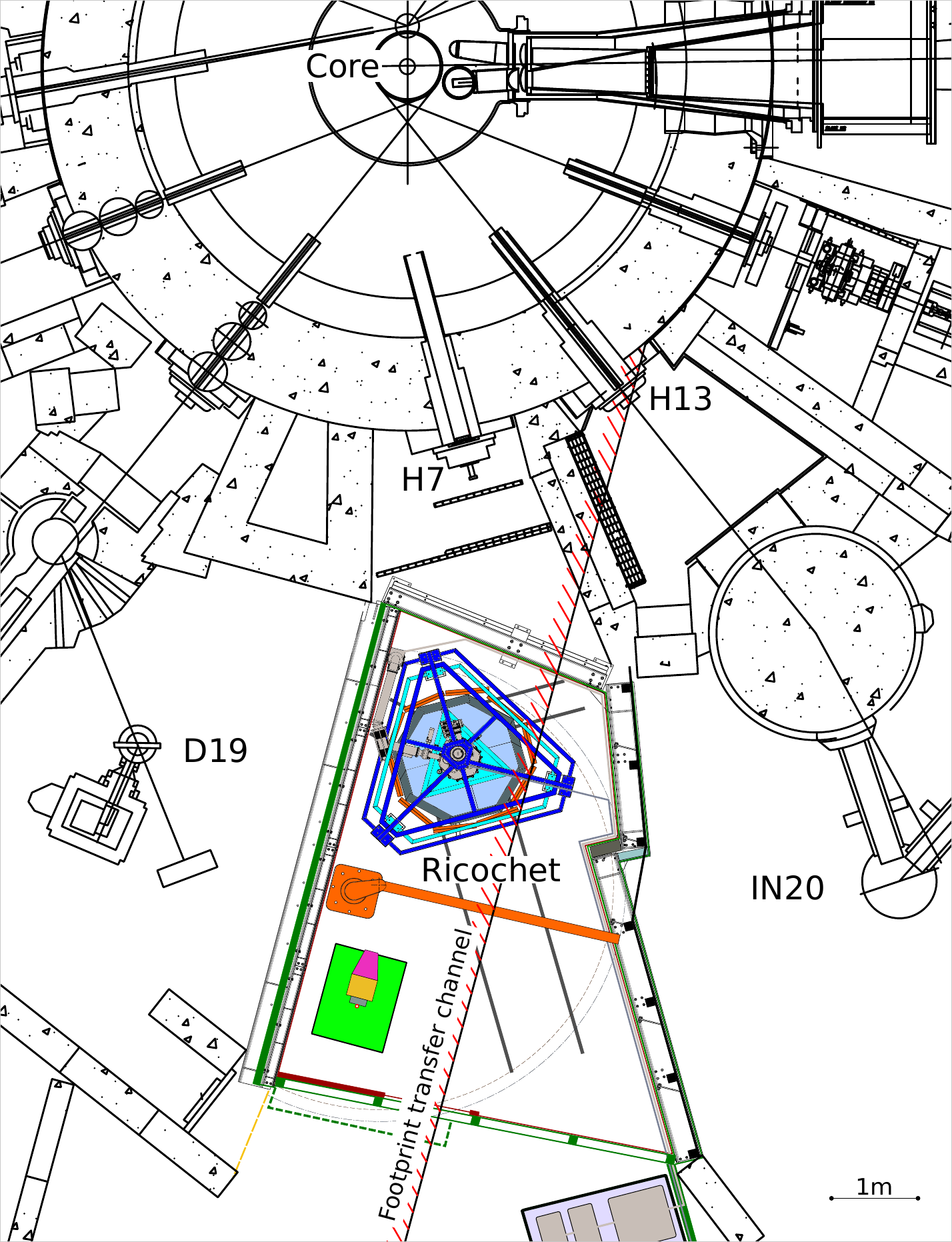}
\includegraphics[width=0.55\textwidth,angle=0]{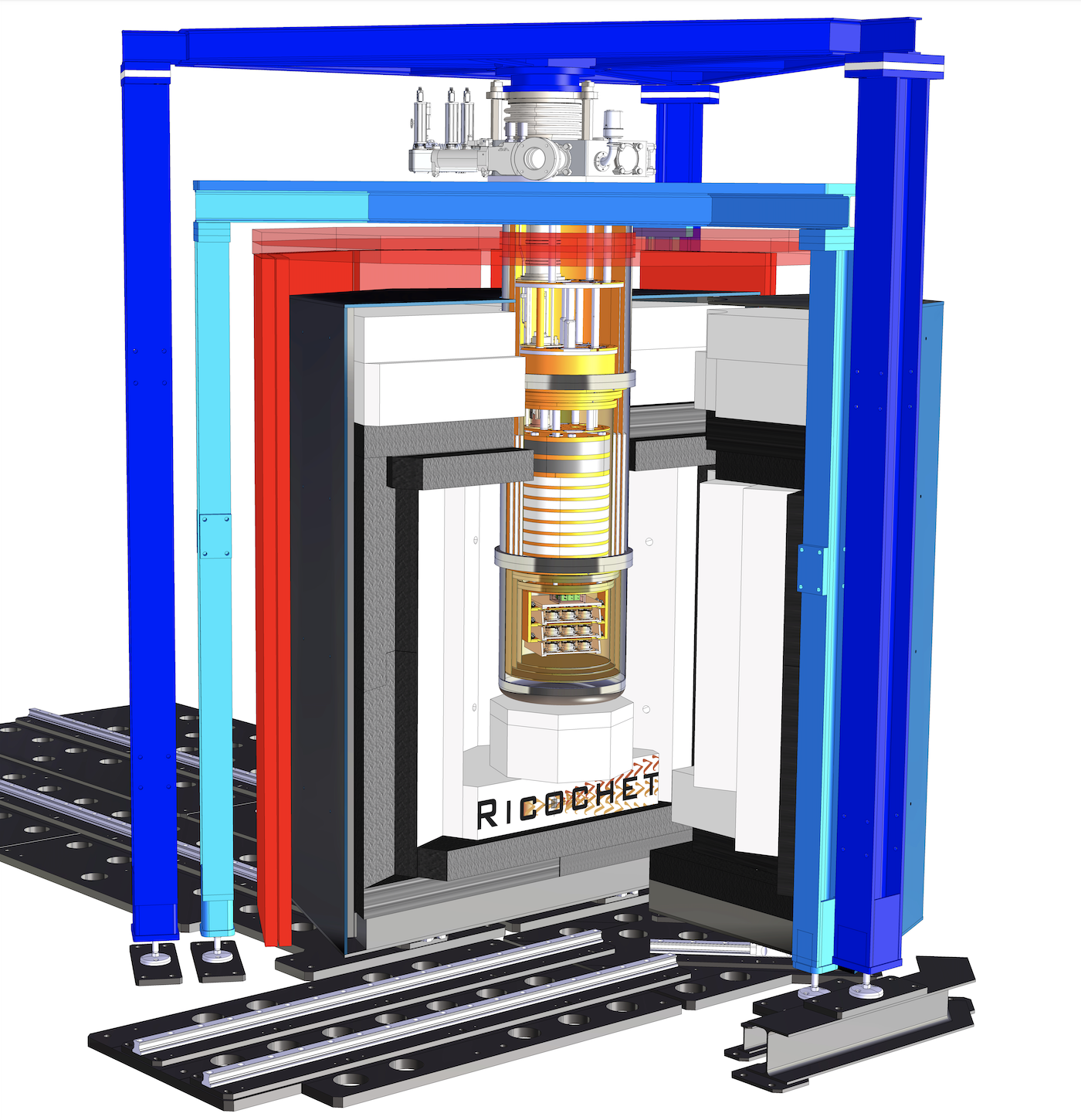}
\caption{{\bf Left:} Schematic of the planned \Ricochet{} integration within the ILL-H7 experimental site. The cryostat is mechanically anchored thanks to two triangle-shaped frames surrounding the passive shielding and active muon veto. Also shown are the 1-t crane (orange), the pulsed DT-based low-energy and mono-energetic neutron source in its storage position (green), and the surrounding IN20 and D19 experiments. {\bf Right:}  Drawing of the future \Ricochet{} experiment. The  Hexa-Dry 200 Ultra-quiet cryostat from CryoConcept is held by two mechanically decoupled frames (dark and light blue) and is surrounded by its outer external shielding layers of polyethylene (white), lead (gray) and soft iron (black). The muon veto is shown as the red panels on the top and side of the setup.} 
\label{fig:RicochetSetup}
\vspace{-0.5cm}
\end{center}
\end{figure*}

The future \Ricochet{} experiment will be deployed at the ILL-H7 site (see Figure~\ref{fig:RicochetSetup}). The H7 site starts at about 7~m from the ILL reactor core that provides a nominal thermal power of 58.3 MW, leading to a neutrino flux at the \Ricochet{} detectors, 8.8 m from the reactor core, of about $1.1\times 10^{12}$~cm$^{-2}$s$^{-1}$ which corresponds to a CENNS event rate of approximately 12.8 and 11.2~events/kg/day with a 50~eV energy threshold and Ge and Zn targets crystal, respectively. The reactor is operated in  cycles  of  typically 50 days duration  with  reactor-off periods  sufficiently  long  to  measure  reactor-independent backgrounds with high statistics, including internal radioactivity and cosmogenic-induced backgrounds. The ILL-H7 experimental site is about 3~m wide, 6~m long and 3.5~m high. It is located below a water channel providing about 15~m.w.e. against cosmic radiation. It is not fed by a neutron beam and is well-shielded against irradiation from the reactor and neighboring instruments (IN20 and D19). Lastly, the operation of the past STEREO neutrino experiment at this site, from 2016 to 2020,  has been successfully demonstrated~\cite{Allemandou:2018vwb}. 

The \Ricochet{} shielding will be divided into two parts: a 300~K outer shielding and a cryogenic inner one.
The outer shielding will be composed of a 35~cm thick layer of 3\%-borated polyethylene to thermalize and capture fast neutrons surrounded by a 20~cm thick layer of lead to mitigate the gamma flux. Additionally, another 35~cm thick layer of polyethylene will be positioned on top to further reduce the cosmogenic fast neutron flux. The whole setup will be surrounded with 0.5~cm thick soft iron to reduce the magnetic stray field originating from neighboring experiments. This outer shielding will be divided into three sections installed on rails to allow for an easy access to the cryostat. Lastly, muon-induced gamma and neutron backgrounds will be further reduced thanks to a surrounding muon veto, made of two layers of 3~cm thick plastic scintillator, to reject events in temporal coincidence with detected muons. The cryogenic inner shielding, installed inside the cryostat above the detectors and composed of a 8.5 cm thick layer of lead and a 21 cm thick layer of polyethylene, with interleaved 1~cm thick copper layers, will ensure a closed shielding. Additionally, 8~mm thick polyethylene layers mounted on each thermal screen will further improve the shielding tightness. Eventually, up to two 1~mm thick layers of mumetal will also be added between thermal screens to further reduce the residual magnetic field from adjacent experiments. Note that the muon veto will also include a cryogenic portion at 50 K to avoid a significant gap in veto coverage at the
crossing of the cryostat. According to our cosmogenic simulations, such a muon veto should exhibit a muon-induced trigger rate of about 400~Hz which will be manageable with our $\sim$100~$\mu$s timing resolution bolometers with a reasonable livetime loss of less than 30\%~\cite{billard:tel-03259707}.

\section{Thermal and fast neutron detection with a low-radioactivity \texorpdfstring{$^3$He}{3He} proportional counter}
\label{sec:He3counter}

 \begin{figure*}[t]
\begin{center}
\includegraphics[width=0.49\textwidth,angle=0]{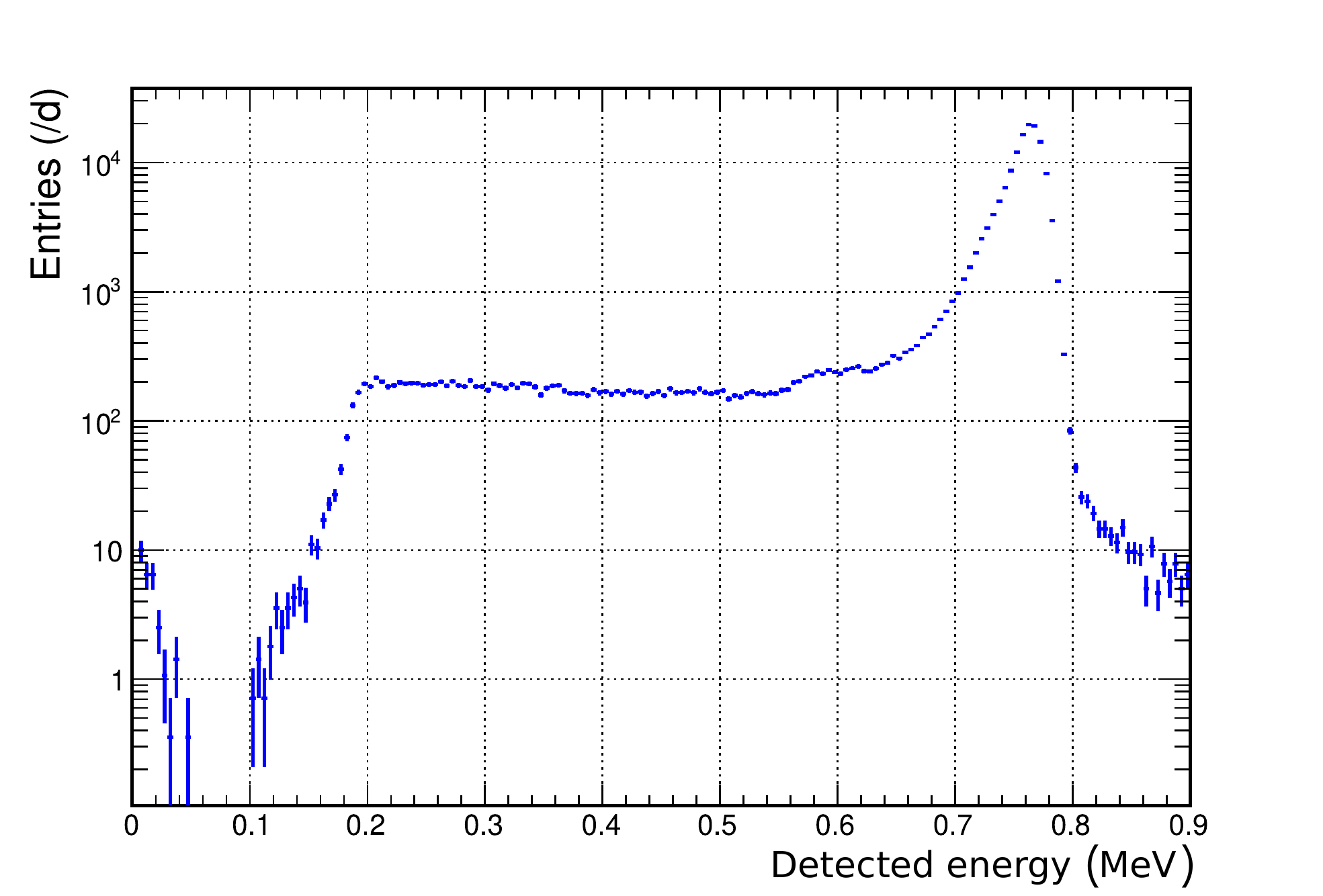}
\includegraphics[width=0.49\textwidth,angle=0]{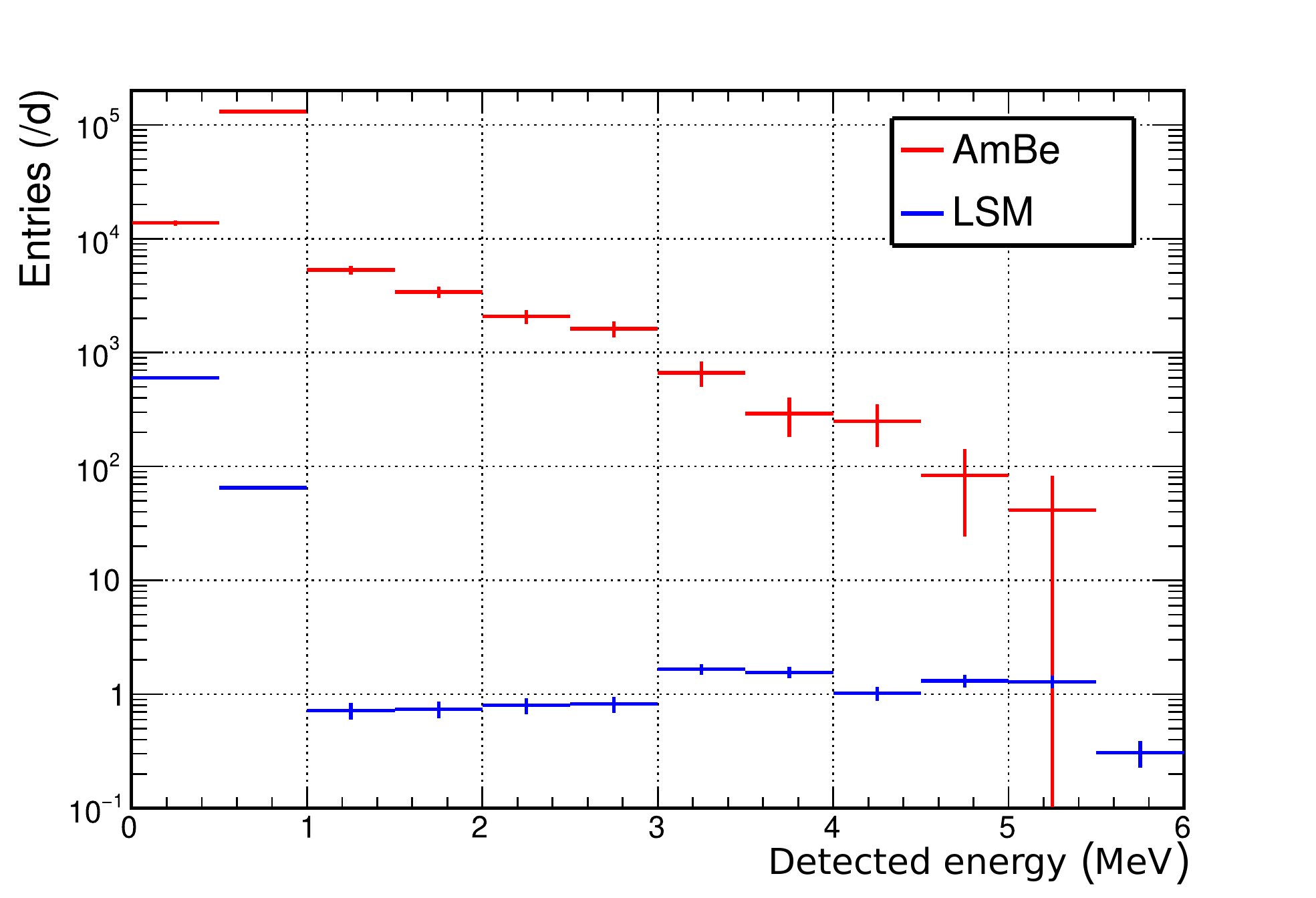}
\caption{{\bf Left:} Measured energy spectrum below 900~keV, covering the so-called thermal neutron capture energy region, taken at the ILL-H7 site without thermal neutron shielding surrounding the $^3$He counter and while the nuclear reactor was in operation. {\bf Right:} Measured energy spectra up to 6~MeV obtained after 49 days of data taking at the Modane underground laboratory (blue) and during a few hours of neutron calibration, using an AmBe source emitting $2\times10^{6}$ neutron per second positioned at one meter from the detector, done at the IP2I (red). Note that events appearing above 1~MeV are expected to be produced by fast neutrons from both elastic scattering, predominantly on $^3$He, and on-flight captures also on $^3$He. In the case of the AmBe neutron calibration a 5~mm thick B$_4$C loaded rubber was surrounding the detector in order to avoid spectral distortions arising from thermal capture pile-up events.} 
\label{fig:ExampleSpectra}
\vspace{-0.5cm}
\end{center}
\end{figure*} 

To characterize the neutron background at the ILL-H7 site, we used a proportional counter tube filled with $^3$He gas. The thermal and fast neutrons are detected via the following on-flight capture reaction:
\begin{equation}
    n \ + \ ^3\text{He} \longrightarrow p \ + \ t \ \ \ (764~\text{keV} \ + \ E{_n})
\end{equation}
where $E_{n}$ is the neutron kinetic energy. The $^3$He(n,p) cross section for thermal neutrons is $\sigma = 5333 \pm 7$ b~\cite{BROWN20181} and drops below several barns for neutron energies between 100~keV and 10~MeV where elastic scattering becomes relevant~\cite{PhysRev.122.1853}. The CHM-57 counter~\cite{Vidyakin} used in this work has an active length of 860 mm with an internal diameter of 31 mm. The counter is filled with 400 kPa of $^3$He and 500 kPa of $^{40}$Ar, where the latter gas element is used as a quencher in order to stabilize the avalanche process of the proportional chamber following an ionization signal detection. Intrinsic backgrounds from alpha decays of U and Th progenies in the walls were reduced by covering the detector's inner walls with 50-60~$\mu$m of Teflon and 1~$\mu$m of electrolytic copper~\cite{Vidyakin}. The ionization signal, predominantly driven by the drifting ions to the external cathode, is read out by an attached Cremat CR-110 single channel charge-sensitive preamplifier. The preamplified signal is then analyzed online by a DT5780 digitizer working in pulse height analysis mode\footnote{For more details see https://www.caen.it/products/dt5780/}.\\

A typical thermal neutron calibration spectrum is shown in Fig.~\ref{fig:ExampleSpectra} (left panel). The expected 764 keV peak from thermal neutron captures is clearly visible.  A broad plateau at lower energies is also seen, resulting from captures occurring near the wall of the counter, where either the triton (t) or proton (p) escapes without depositing its full energy. From Fig.~\ref{fig:ExampleSpectra} (left panel) two shoulder-like structures, due to this so-called wall effect, are clearly visible at 191~keV and 563~keV which respectively correspond to the full collection of only the triton or proton recoils. These three characteristic features in the energy spectrum, at 191~keV, 563~keV and 764~keV, have been used to cross-check the energy scale and linearity of the detector response~\cite{Vidyakin}. According to SRIM-based simulations~\cite{SRIM}, and further confirmed with our Geant4 simulations detailed in Sec.~\ref{sec:PCsimulation}, the averaged proton and triton track lengths following a thermal neutron capture on $^3$He are about 2~mm and 0.7~mm, respectively.

 \begin{figure*}[t]
\begin{center}
\includegraphics[width=0.51\textwidth,angle=0]{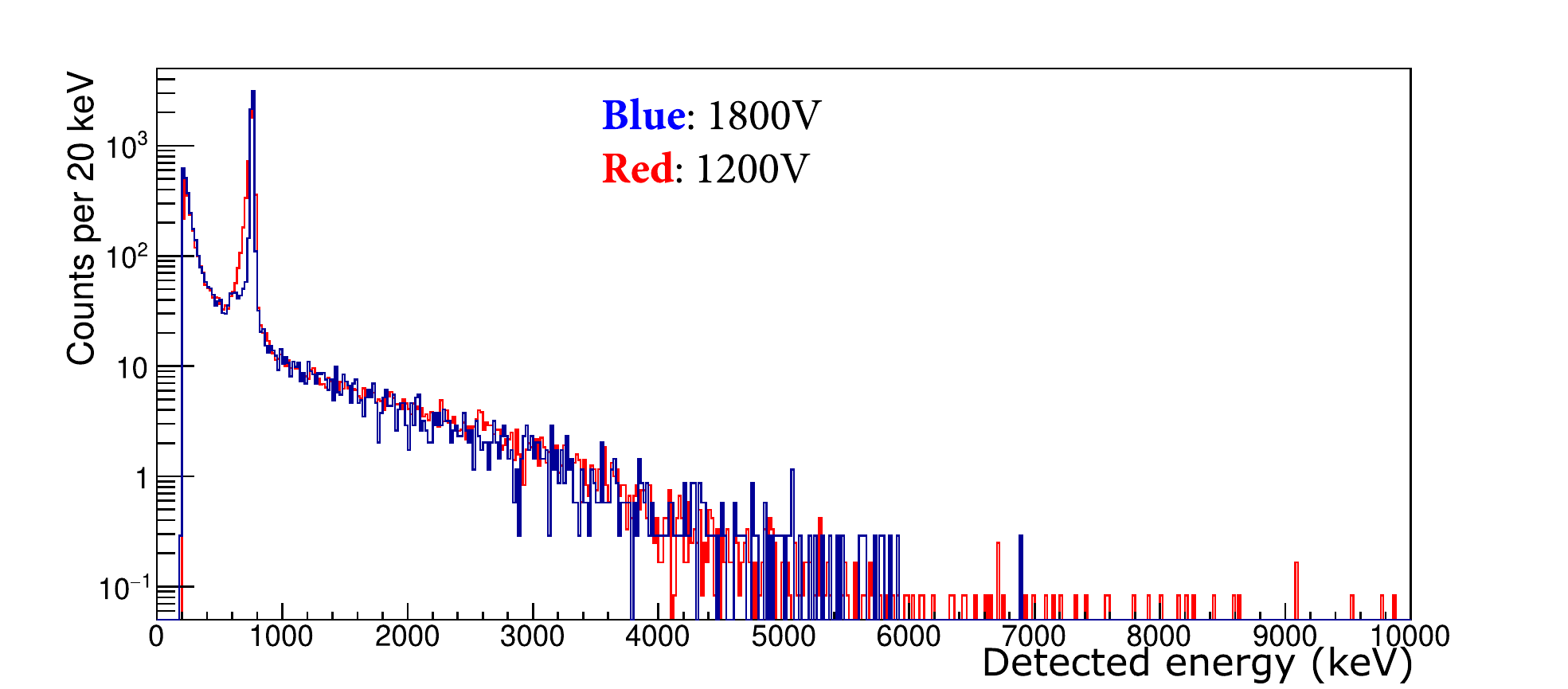}
\includegraphics[width=0.47\textwidth,angle=0]{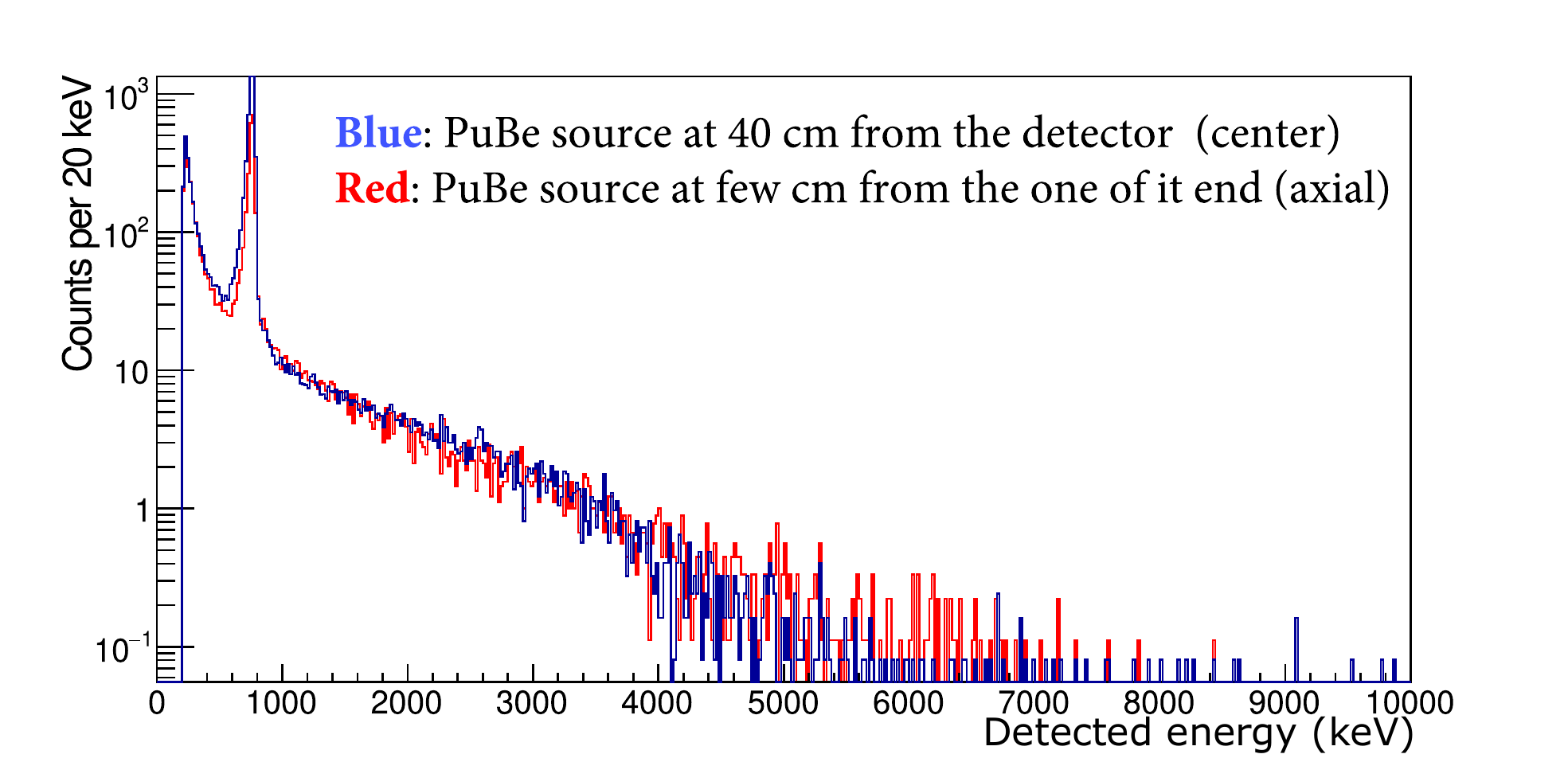}
\caption{{\bf Left:} Observed energy spectra with the $^3$He counter irradiated by a PuBe neutron calibration source with two different amplification voltages of 1200~V (red) and 1800~V (blue) for which the avalanche amplification gains differ by a factor of about 6. {\bf Right:} Same as left panel but with a fixed voltage of 1200~V and the neutron source either irradiating the counter along its axial (red) or radial axis (blue).} 
\label{fig:DubnaTests}
\vspace{-0.5cm}
\end{center}
\end{figure*}

In this work, we focus on the high energy portion of the observed energy spectrum, {\it i.e.} above 1~MeV in detected energy, in order to estimate the fast component of the neutron background at the ILL reactor. For energies beyond 1~MeV, we expect events to be predominantly due to elastic and inelastic (on-flight captures) scatterings of fast neutrons on $^3$He. Note that elastic scatterings on $^{40}$Ar nuclei are expected to have a negligible contribution to the observed energy spectrum beyond 1~MeV as these would require neutron energies above 20~MeV due to both kinematics and their 50\%-60\% ionization yield at a few MeV in recoiling energy (see Sec.~\ref{sec:PCsimulation}). Also, thanks to their much lower stopping power, gamma induced electronic recoils cannot deposit more than a few hundreds of keV in the detector volume. Eventually, the only relevant background beyond 1~MeV of detected energy is coming from alpha decays with degraded energies arising from residual radioactive contaminants. As mentioned above, this $^3$He proportional counter has been designed to minimize such  contamination in order to offer a maximal sensitivity to fast neutron detection. 

Figure~\ref{fig:ExampleSpectra} (right panel) shows two observed energy spectra obtained with this counter when it was irradiated by a fast neutron source of AmBe (red) and when it was operated in the low-background Modane underground laboratory (LSM)~\cite{Lemrani:2006dq} (blue). Focusing on the energy range above 1~MeV, where we expect to detect fast neutrons, we see a clear excess of events during the AmBe calibration with respect to the low-background measurement performed at LSM. Based on previous neutron measurements done at LSM~\cite{Rozov:2010bk}, the observed events beyond 1~MeV are understood as residual radon contamination, resulting in a flat background of 2 events/day/MeV that will ultimately limit our neutron detection sensitivity, see Sec.~\ref{sec:ValidationIP2I}.\\

In order to validate our approach of using the observed energy spectrum above 1~MeV to estimate the fast neutron background, we performed two additional cross checks of the detector response dedicated to the linearity of the energy scale and its sensitivity to the incoming neutron direction. As the deposited energy increases, one can expect to observe so-called space charge effects corresponding to a degradation of the amplification gain due to charge screening~\cite{Leder:2017lva}. The latter is directly related to the amplification gain, such that a larger gain would lead to higher charge screening due to a larger number of electrons produced in the avalanche process. Figure~\ref{fig:DubnaTests} (left panel) shows two measurements where the voltage was varied from 1200~V to 1800~V, corresponding to an amplification gain variation of about 6. Because we observe no statistically significant change in the spectrum under this widely varied gain, the 1650 V operating voltage is taken to be in the linear regime, at least for our region of interest up to 10 MeV. Note that variations of the ionization yield as a function of the recoil energy of $^3$He, proton, and triton could also lead to non-linearity in the energy-scale. However, SRIM simulations of all three nuclei from 500~keV up to 10~MeV of recoiling energy, in 400 kPa of $^3$He and 500 kPa of $^{40}$Ar gas, predict an ionization yield between 98.3\% and 100\% with negligible energy dependence (see Sec.~\ref{sec:PCsimulation}). Additionally, these simulated ionization yield results are further supported by the experimental observation from \cite{PhysRev.122.1853} where a similar $^3$He-based proportional counter and mono-energetic neutrons with energies up to 17.5~MeV were used and no significant variations in the ionization yields of p, t, and $^3$He was found.

 \begin{figure*}[t]
\begin{center}
\includegraphics[width=\columnwidth,angle=0]{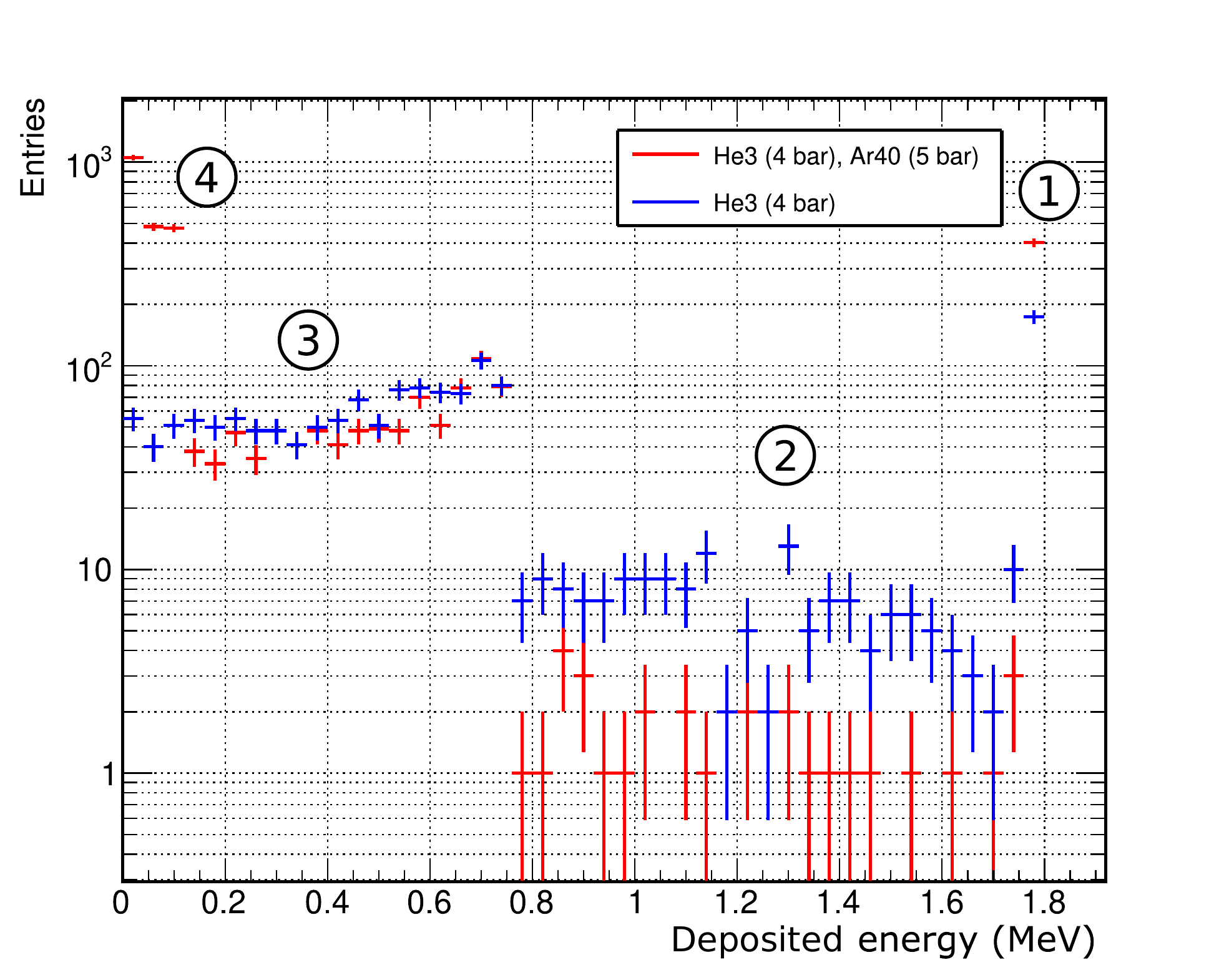}
\includegraphics[width=\columnwidth,angle=0]{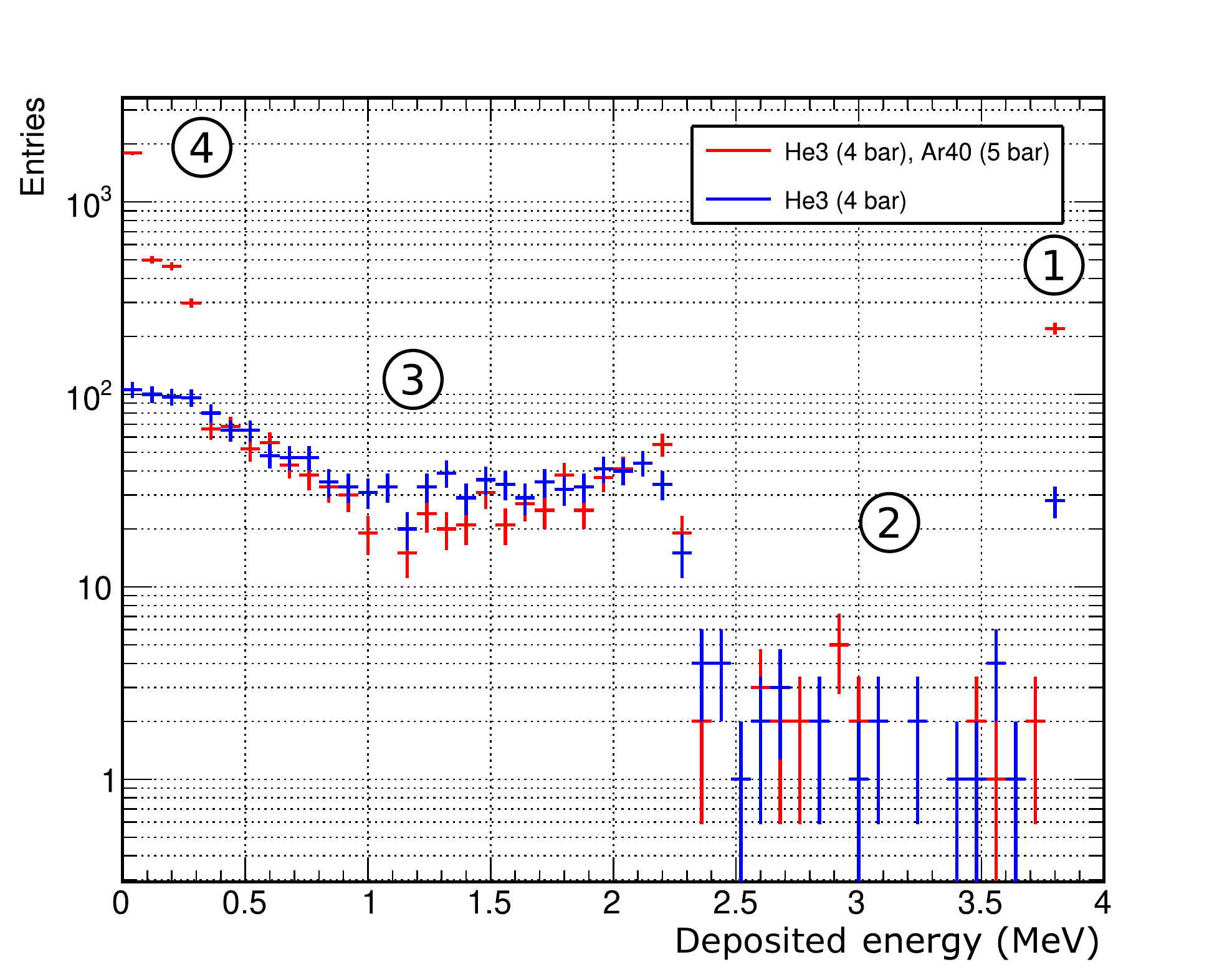}
\caption{Simulations of the $^3$He proportional counter response to an isotropic mono-energetic neutron flux of 1~MeV ({\bf left}) and 3~MeV ({\bf right}). The resulting spectra are shown in two configurations: a pure $^3$He gas at 400 kPa (blue) and the gas mixture corresponding to our detector gas composition made of 400 kPa of $^3$He + 500 kPa of $^{40}$Ar (red).} 
\label{fig:SimulatedHe3}
\vspace{-0.5cm}
\end{center}
\end{figure*} 

The fast neutron flux is expected to be anisotropic at the ILL-H7 reactor site and several localised sources have been identified in previous measurements done by the STEREO collaboration~\cite{Allemandou:2018vwb}. Therefore, we investigated the response of our detector to a neutron calibration source irradiating our detector in two extreme orientations: centered along its z axis with a radial orientation, and positioned at the bottom endcap of the detector offering an axial orientation. The resulting energy spectra are shown in Fig.~\ref{fig:DubnaTests} (right panel) for the radial (blue) and axial (red) neutron source irradiation orientations. From the comparison of these two extreme cases, we only observe a marginal difference at the highest energies, {\it i.e.} above 6~MeV. This is explained by the improved full collection efficiency of the recoiling nuclei when their tracks are aligned with the detector axis.

Based on these results, we can conclude that our $^3$He proportional counter is well-suited to measure and characterize the fast neutron component of the ILL-H7 reactor site where the future \Ricochet{} experiment will be deployed.

\section{Geant4 simulations}

The goal of this work is to compare our observed energy spectra to simulated ones in various conditions and from different sites, both in terms of shape and rate. Therefore, in the following section we discuss the details of our simulations. Those include both the simulation of the $^3$He counter response and of the different cosmic and reactor neutron sources. All of the following simulations have been done within the Geant4 10.06.p02 software considering the ``Shielding'' physics list~\cite{Allison:2016lfl}.

\subsection{\texorpdfstring{$^3$He}{3He} proportional counter simulation}
\label{sec:PCsimulation}

The $^3$He proportional counter is simulated according to its geometry and gas composition as described previously. Based on our observed  $\sim$30~keV energy resolution (RMS), far smaller than the considered bin width of 250~keV when compared to our measured spectra, and the negligible space charge effect, we did not include these finite detector response effects in our simulations. However, note that the physics list incorporates the “G4ScreenedNuclearRecoil” module that models screened electromagnetic nuclear elastic scattering, as required for an accurate simulation of the propagation of the proton and triton after a neutron capture on $^3$He or following any elastic scattering happening in the proportional counter~\cite{Mendenhall_2005}. Lastly, using SRIM-based recoil simulations of proton, triton, $^3$He and Ar from 500~keV up to 10~MeV, we found the ionization yield of the three lighter nuclides to be greater than 98.3\% (at 500 keV) and rising up to almost 100\% at 10~MeV. For the Ar recoils however, we found the ionization yield to be of 48.5\% at 500~keV and constantly rising up to 93.5\% at 10~MeV~\cite{SRIM}. Taking into account recoil kinematics and a 1~MeV energy threshold in detected energy, we expect to only be sensitive to p, t, and $^3$He recoils for which we can assume that the ionization yield is constant and that the detected energy is equivalent to the kinetic recoil energy.

Geant4 simulations were performed in which monoenergetic neutron fluxes were isotropically incident on the $^3$He proportional counter.  Figure~\ref{fig:SimulatedHe3} presents the resulting energy spectra for incident neutron energies of 1 MeV (left panel) and 3 MeV (right panel). Both panels present the results with two gas compositions: pure $^3$He gas at 400~kPa (blue) and the actual gas mixture of our detector made of 400~kPa of $^3$He and 500~kPa of $^{40}$Ar (red). For both panels we see four characteristic features: 1) a line at $E_n + 764$~keV corresponding to on-flight neutron captures fully collected in the detector volume, 2) on-flight neutron captures happening near the wall of the detector with lowered energies deposited inside the gas, 3) a rather flat $^3$He recoil energy spectrum with its corresponding endpoint at $\frac{3}{4}E_n$, and 4) a low-energy $^{40}$Ar recoil energy spectrum contribution with its expected endpoint at $0.1\times E_n$ (when $^{40}$Ar gas is added to the mixture). Interestingly we see that the addition of the 500~kPa of $^{40}$Ar gas has very little effect on the observed energy spectrum of the $^3$He recoils but  has the benefit of increasing the peak-to-continuum ratio of on-flight neutron captures, hence improving the spectroscopic ability of the detector. This is explained by the fact that this additional gas component increases the fraction of fully collected proton + triton tracks by increasing the pressure hence reducing the recoiling nuclei track lengths. As a conclusion of these simulations, we expect our proportional counter to exhibit some neutron spectroscopic capabilities ({\it i.e.} direct neutron energy measurement) even though these are attenuated by the $^3$He recoil contributions from neutron elastic scatterings and by incomplete track collections. In spite of these limitations, Fig.~\ref{fig:SimulatedHe3} illustrates the capability of our detector to assess the fast neutron flux at the \Ricochet{} experiment, both in energy dependence and magnitude (see Sec.~\ref{sec:NeutronFluxMeasure}).

\subsection{Cosmogenic and reactogenic neutrons}
\label{sec:simuneutronflux}
The \Ricochet{} experiment will be using low-radioactivity materials such that the internal radioactivity is expected to be sub-dominant with respect to the external cosmogenic and reactogenic neutrons. 
To simulate the cosmogenic neutrons at the various sites of interest, we used the Cosmic-ray shower library (CRY) that generates correlated cosmic-ray particle shower distributions for use as input to our Geant4 transport and detector simulation codes~\cite{CRY}. We considered the latitudes of Grenoble for the ILL-based simulations and of Lyon for the IP2I-based simulations that are relevant for the geomagnetic cut-off. Additionally, the live-time simulated by CRY with its otherwise default settings has been divided by 1.28 for the ILL site, as suggested by past muon flux measurements at the ILL-H7 site~\cite{Allemandou:2018vwb}.

 \begin{figure}[t]
\begin{center}
\includegraphics[width=\columnwidth,angle=0]{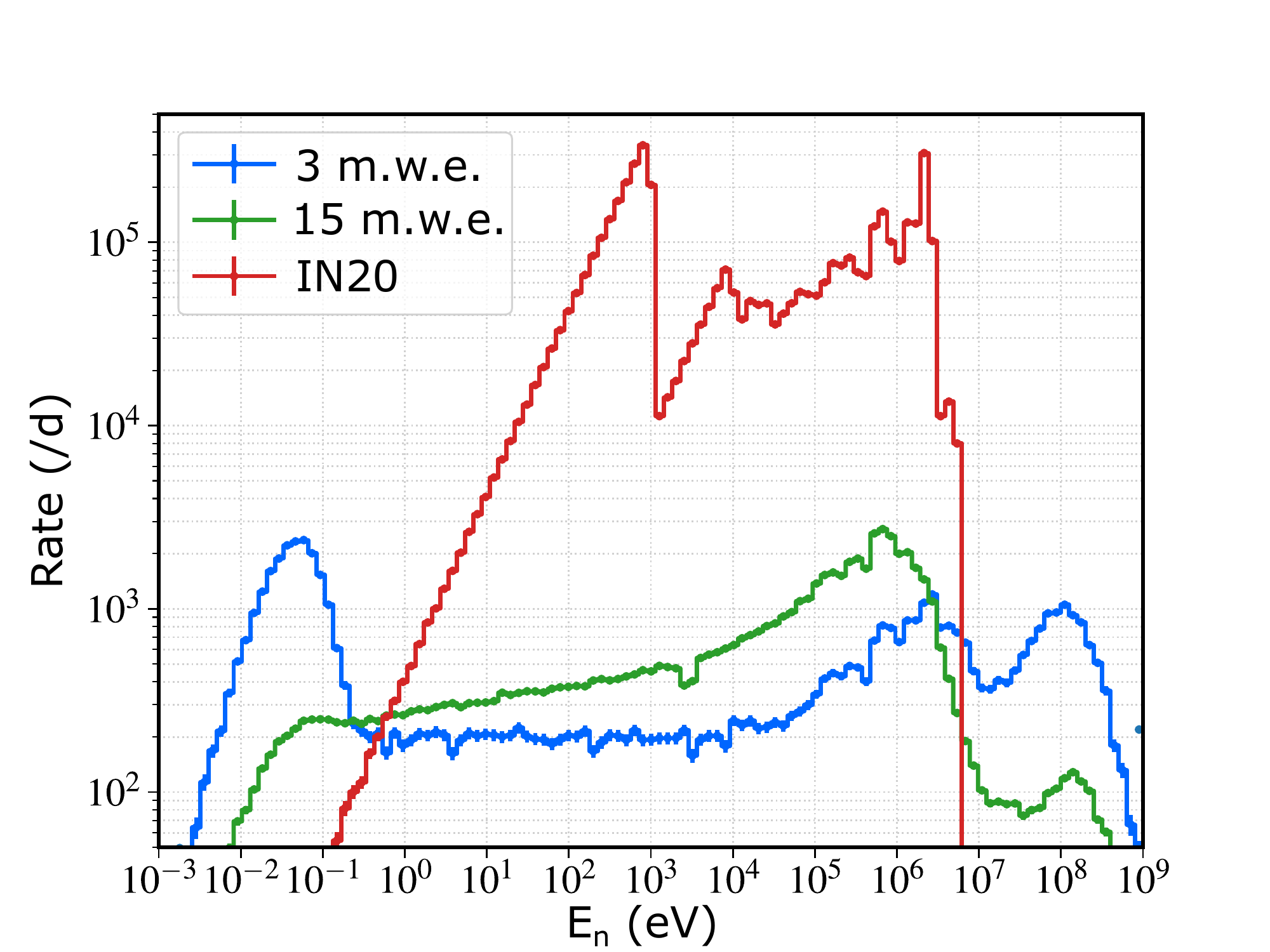}
\caption{Simulated neutron fluxes entering the $^3$He detector per day as a function of energy for the reactogenic (IN20 - red from~\cite{pequignot:tel-01217946}), and the cosmogenic neutrons component at the two different locations considered in this work: IP2I surface lab with a $\sim$3~m.w.e. of overburden (blue), and the ILL-H7 reactor site with its mean 15 m.w.e. of overburden (green). Note that the binning along the x-axis is logarithmic. } 
\label{fig:He3Fluxes}
\vspace{-0.5cm}
\end{center}
\end{figure}

Concerning the reactogenic neutrons at ILL, we used simulations performed by the STEREO collaboration. From the background measurements done in preparation to the STEREO experiment~\cite{Allemandou:2018vwb}, the main source of reactogenic background identified was the IN20 experiment and, more specifically, the corresponding neutron beam H13 and its shutters. Using a MCNP code, the reactor neutron energy spectrum has been propagated through the H13 tube and the IN20 experimental site to estimate the energy spectrum and rates at the STEREO location~\cite{pequignot:tel-01217946}. However, the geometry did not include some shielding walls that were added since. Therefore, we expect the energy spectrum to be overestimated and we consider it as a conservative upper limit. The overall normalization of the flux is 790 neutrons/m$^2$/s at reactor nominal power.

Figure~\ref{fig:He3Fluxes} shows the simulated reactogenic and cosmogenic neutron spectra entering the $^3$He proportional counter. The reactogenic spectrum was obtained at 58~MW nominal thermal power for a box-like generation surface of 56~m$^2$ (IN20 - red). The cosmogenic spectra are from two different locations: IP2I surface lab with its averaged overburden of $\sim$3~m.w.e. (see Sec.~\ref{fig:ResultsIP2I} - blue), and the ILL-H7 reactor site with its mean 15 m.w.e. of overburden (green)~\cite{Allemandou:2018vwb}. In the IP2I surface lab case we can clearly identify the four usual cosmic neutron populations: thermal ($E_n < 0.5$~eV), epithermal ($0.5~\text{eV} < E_n < 0.1~\text{MeV}$), evaporation ($0.1~\text{MeV} < E_n < 20~\text{MeV}$), and cascade ($E_n > 20~\text{MeV}$). However, when considering the ILL-H7 site, and its averaged artificial overburden of 15 m.w.e. (see Sec.~\ref{sec:ILLneutronMeasure}), we see that most thermal and cascade neutrons are cut-out and that the evaporation neutron population has shifted to lower energies with its peak at around 1~MeV. Though significantly reduced with respect to an unshielded surface lab, we still observe some high energy neutrons up to 200~MeV that can still affect the future \Ricochet{} experiment sensitivity. Regarding the reactogenic IN20 model (red histogram), we see that its MeV-scale neutron flux is more than one order of magnitude larger than its cosmogenic counter part (green histogram), but it also exhibits a much lower energy end point of 6~MeV suggesting that it should be better attenuated by the \Ricochet{} shielding.

 \begin{figure*}[t]
\begin{center}
\includegraphics[width=0.8\textwidth,angle=0]{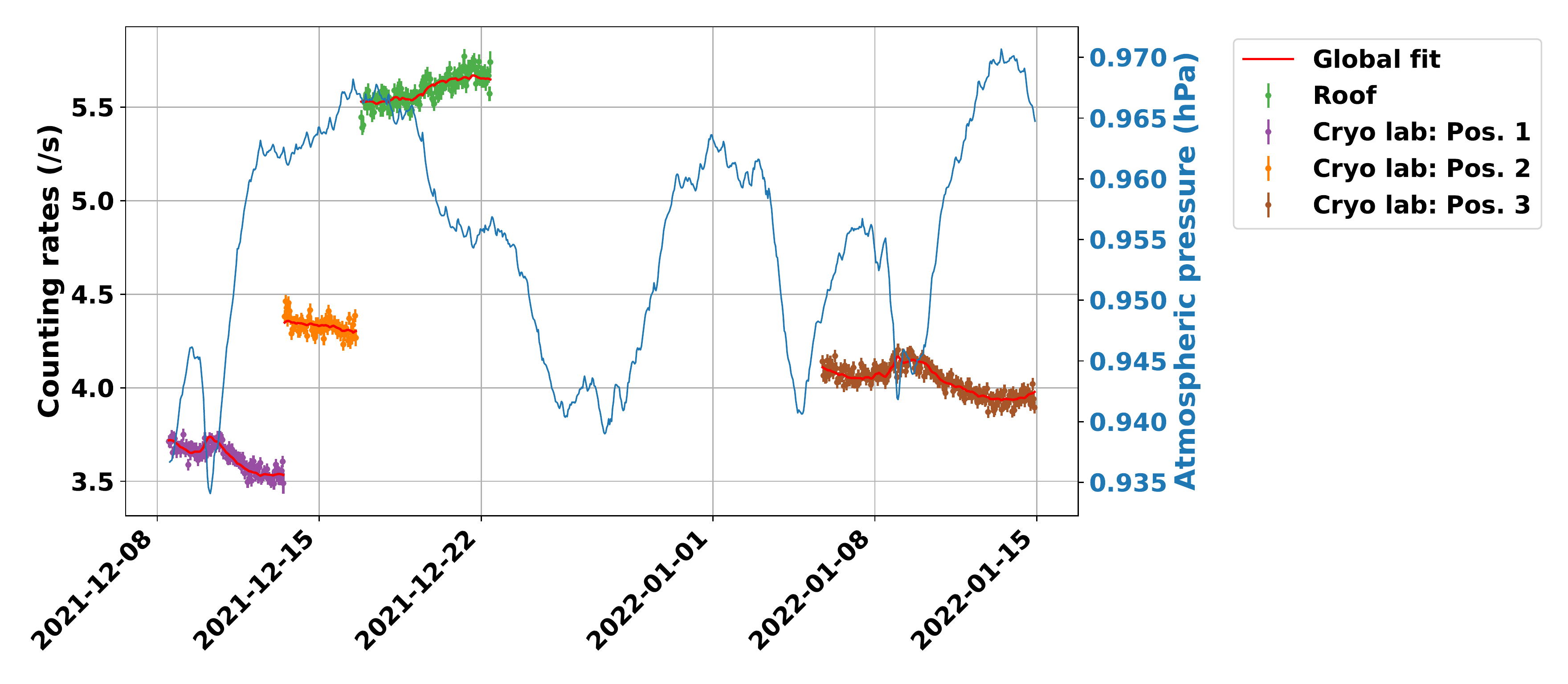}
\caption{Time evolution of the muon trigger rate at three different locations in our IP2I cryogenic lab (purple, orange, and brown dots) and from the roof of the building to determine the muon trigger rate with no overburden (green dots). The blue curve shows the  evolution in time of the atmospheric pressure as extracted from the ERA5 global reanalysis hourly data~\cite{ERA5}. The latter was taken into account in our determination of the mean muon trigger rates at each locations within our fitting procedure (red line).} 
\label{fig:muon}
\vspace{-0.5cm}
\end{center}
\end{figure*}

\section{Fast neutron flux characterizations}
\label{sec:NeutronFluxMeasure}

This section is the core of our work as it discusses how our simulated neutron backgrounds compare with our experimental observations with the $^3$He proportional counter presented in Sec.~\ref{sec:He3counter}. It is worth emphasizing that no parameter of the reactogenic and cosmogenic neutron flux models was tuned to better reproduce the observed spectra. Therefore, both neutron flux models have been used as is to compute our expected \Ricochet{} neutron background presented in Sec.~\ref{sec:RicochetBackground}.

\subsection{Validation of the method: the IP2I fast neutron background}
\label{sec:ValidationIP2I}
As a proof of concept of our proposed neutron background assessment methodology we first studied the case of the IP2I surface lab. The latter is located in Lyon at an altitude of 181 meters above sea level and at a latitude of 45$^\circ$45'32.616'' North. The modelization of the cryogenic lab in our CRY simulations considers that it is in the basement of a two-story high building made of thick concrete walls and floors. We found that the main overburden comes from the floor and ceiling above our experimental area, which amounts to 1.2~m of concrete and consequently provides about 2.76~m.w.e. of direct vertical overburden. Additionally, the near proximity of our detectors to a 1.45~m-thick concrete wall provides an additional position dependent solid angle-integrated overburden.

In order to properly compare our cosmogenic simulations to our observations with both the $^3$He proportional counter and the Ge bolometers operated in the same lab, about 3 meters away from each other, we first estimated the common overburden with the use of muon flux attenuation measurements. To do so, we used 1~cm thick, 20~cm long, and 5~cm wide plastic scintillator panels arranged in a 4$\times$4 array from the DIAPHANE  experiment~\cite{Marteau:2016jcn}. The energy loss from the muons going through the panels is converted into scintillation photons which are guided towards a multi-anode photomultiplier by    wavelength-shifting optical fibres. Muons were identified as such by requiring coincident triggers on all four plastic scintillators planes. 
%Because of the non-uniform overburden provided by the building above our cryogenic lab, we decided to measure the muon rates at three different locations to derive an averaged overburden to be considered in following analyses. 
In order to confirm the IP2I building geometry utilized in the
simulations for the neutron background assessment, we measured the muon rates at three different locations and derived an averaged overburden.
The first position was next to the $^3$He counter but closer to the thick wall (maximizing the effective overburden). The second position was three meters away against the opposing thin wall next to the windows (minimizing the effective overburden). Lastly, the third position was above the cryostat where the Ge detectors were operated. Therefore, the latter position is the most relevant while the first two ones can be considered as being the upper and lower bounds on the surface lab overburden.
%We do not know precisely the muon detection efficiency of our detector which is expected to be around 90\% based on the observed signal amplitude spectra of each scintillators. Note however that as we are only doing relative measurements here to estimate the overburden of the lab from attenuation measurements, knowing precisely this efficiency is not required. 

 \begin{figure*}
\begin{center}
\includegraphics[width=0.43\textwidth,angle=0]{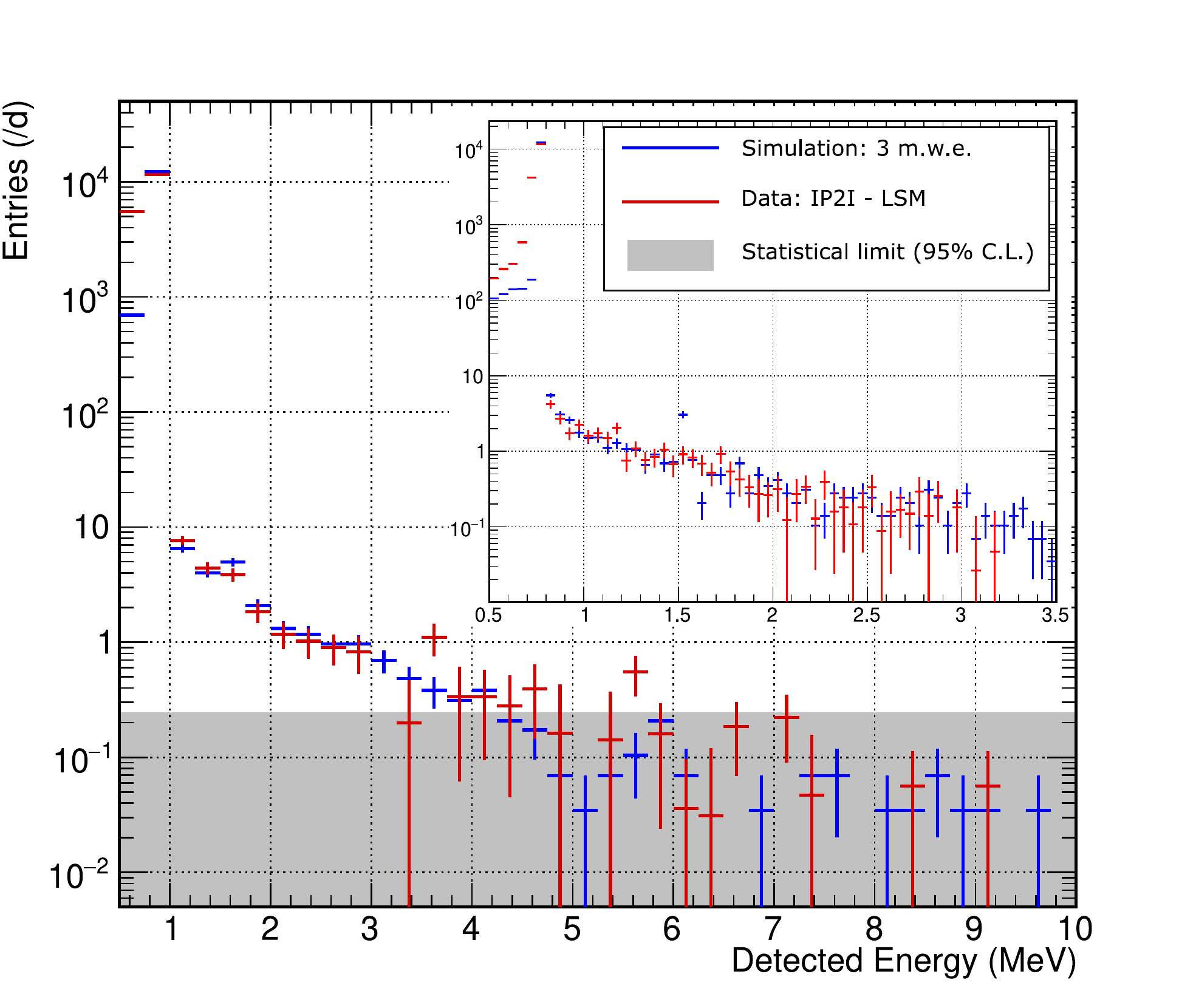}
\includegraphics[width=0.53\textwidth,angle=0]{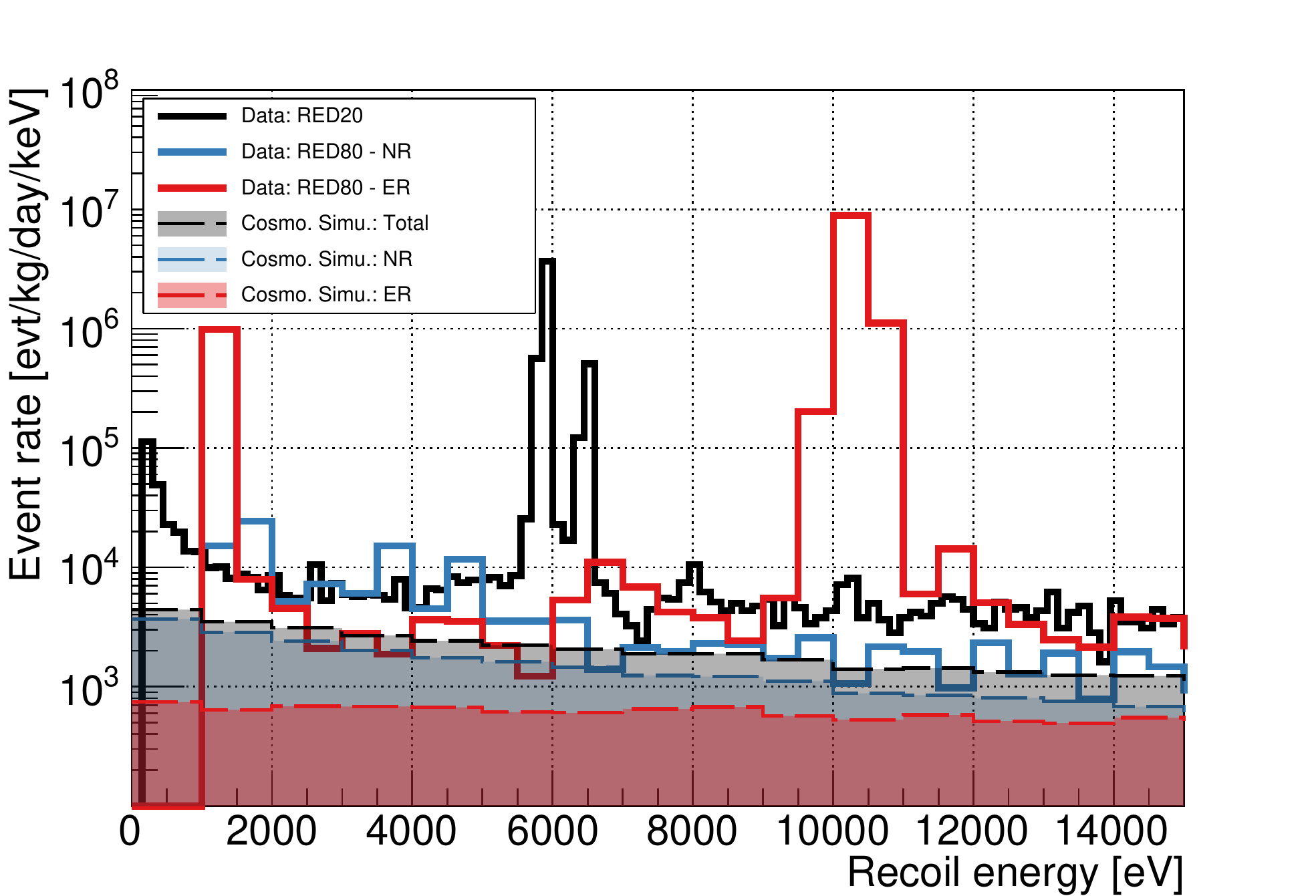}
\caption{{\bf Left:} Comparison of the simulated (blue) and the observed (red) $^3$He proportional counter data while operated at the IP2I surface lab after 18 days of data taking. The red histogram has been obtained by subtracting the observed spectra from the LSM to subtract the internal alpha background of the detector. The simulation considers the IP2I building geometry discussed in the text with its averaged overburden of 3~m.w.e. as estimated from our muon flux attenuation measurements. As discussed in Sec.~\ref{sec:He3counter}, due to the internal proportional counter background, a statistically significant neutron contribution can be inferred from an observed rate larger than about 0.25~evt/day/bin at 95\% C.L. {\bf Right:} Comparison between the observed data with the RED20~\cite{Armengaud:2019kfj} and RED80~\cite{misiak:tel-03328713} low-threshold cryogenic detectors, operated in the IP2I cryostat surrounded by a 70\% coverage 10~cm thick lead shielding, and its resulting cosmogenic background simulation. For both the simulation and the RED80 data, we show the nuclear (blue) and electronic (red) recoil components, as well as the total expected cosmogenic background at IP2I (black dashed curve). Note that the remaining internal and external radioactivity from the surrounding materials are not taken into account here. For the sake of clarity, error bars are not shown but are about 10\%~\cite{Armengaud:2019kfj} and 30\%~\cite{misiak:tel-03328713} for the RED20 and RED80 data, respectively.} 
\label{fig:ResultsIP2I}
\vspace{-0.5cm}
\end{center}
\end{figure*}

Figure~\ref{fig:muon} shows the time evolution of the observed muon trigger rates at these three locations within our cryogenic lab (purple, orange, and brown dots) and from the roof of the building (green dots) to determine a zero-overburden reference measurement. Also shown is the time dependent atmospheric pressure (blue line) which was used in our fitting model (red line) to derive a mean muon trigger rate at each location. Thanks to the muon trigger rate from the roof, we can derive the muon flux attenuations $a_\mu$ at the three cryogenic lab locations which were found to be of: $0.63\pm 0.01$ (position 1), $0.78\pm 0.01$ (position 2), and $0.72\pm 0.01$ (position 3). Following the procedure described in~\cite{Angloher:2019flc}, corresponding overburdens $m_0$ can be estimated from the observed muon flux attenuation factors $a_\mu$ using the approximation below from~\cite{Theodorsson}:
\begin{equation}
    a_\mu = 10^{-1.32\log d - 0.26 (\log d)^2}
\end{equation}
where $d = 1 + m_0/10$, and $m_0$ is given in meter water equivalent (m.w.e.). The derived overburden values at each of these locations were thus found to be: $4.05 \pm  0.16$~m.w.e. (position 1), $2.04 \pm  0.11$~m.w.e. (position 2), and $2.76 \pm  0.13$~m.w.e. (position 3), leading to an averaged overburden in our lab considered hereafter of $2.95 \pm  0.65$ m.w.e. 
%Interestingly, the muon flux attenuation obtained from our CRY simulations of the IP2I cryogenic lab was found to be 0.65, hence compatible with our observed data.
Interestingly, attenuations obtained from our CRY simulations of the muon panel setups at position 1 and 2 of the I2PI lab were found to be 0.65 and 0.78, respectively, which supports the IP2I geometry used hereafter.

Figure~\ref{fig:ResultsIP2I} (left panel) shows the comparison between the observed and simulated $^3$He spectra obtained at the IP2I surface lab. The measured energy spectrum (red histogram) has been obtained by subtracting the observed event rate from LSM in order to remove the internal background of the detector (see Sec.~\ref{sec:He3counter}). The corresponding observed fast neutron rate with detected energies greater than 1~MeV, is of $25.6 \pm 1.5$ per day. As one can conclude from Fig.~\ref{fig:ResultsIP2I} (left panel), the observed and simulated spectra match almost perfectly well over the entire energy range relevant for fast neutron flux measurements ({\it i.e.} for detected energies above 1~MeV). This suggests that both the magnitude and energy dependence of the fast neutron flux entering the $^3$He proportional counter is well estimated by our simulations up to about 4.5~MeV in detected energy -- limited by the $^3$He proportional counter's internal background subtraction limit shown as the gray contour. The latter represents the 95\%~C.L. limit on the significance of the neutron detection rate, calculated using the impact of Poisson fluctuations on the internal background subtraction described in Sec.~\ref{sec:He3counter}.\\

In order to further validate this cosmogenic neutron flux model, we propagated it to 38~g Ge cryogenic bolometers operated in a dry dilution cryostat surrounded by a 70\%-coverage 10~cm thick cylindrical lead shielding with a 7~cm thick bottom end-cap. Figure~\ref{fig:ResultsIP2I} (right panel) shows the comparison between the observed recoil energy spectra from our prototype bolometers called RED20~\cite{Armengaud:2019kfj} (black solid line) and from RED80, which has the ability to discriminate electronic recoils (red) from nuclear ones (blue)~\cite{misiak:tel-03328713}, and the simulated cosmogenic background (filled histograms). Note that our simulations do not take into account internal and external radioactivity from the surrounding materials which are likely to also contribute to the total background, especially with an incomplete lead shielding as considered here. Also, the cryogenic Ge bolometers were calibrated using a $^{55}$Fe source emitting 5.89 and 6.49~keV x-rays for RED20, and internal $^{71}$Ge electron-capture decays emitting  low-energy x-rays of 10.37 and 1.3~keV following a thermal neutron activation of the RED80 detector. Overall, from 1 to 15~keV we see that the total observed and simulated recoil spectra agree within a factor of about three\footnote{The steep rise in the energy spectrum below 1~keV, so-called low-energy excess, is the subject of ongoing intense worldwide investigations. For more details, see~\cite{Proceedings:2022hmu}. Additionally, note that the sharp rise at 1.5~keV in RED80 is due to the 1.3~keV x-ray line from $^{71}$Ge electron-capture decays}. Thanks to RED80, which benefits from particle identification capability with its double heat-and-ionization readout, we see that this disagreement is about a factor of six for the gammas and three for the neutrons. However, it is worth noticing that the simulation reproduces well the different slopes of the observed electronic and nuclear recoil spectra. The gamma discrepancy is most likely explained by an underestimation of the gamma background in our cosmogenic-only simulations where radiogenic contributions are not taken into account while they are likely to be significant. Indeed, removing the lead shielding around the cryostat increases the electronic recoil rate in the bolometers by a factor ten, while a more optimized shielding should provide order of magnitude better protection from gamma rays~\cite{Heusser:2015ifa}. The observed excess of the electronic recoil rate compared to a simulation restricted to cosmogenic gammas is thus not surprising. The factor of three discrepancy between the simulations and the observations regarding the neutron component is however still under investigation. For instance, some plausible explanations could be that our IP2I experimental setup simulation is oversimplified, or that our cryogenic lab exhibits a larger than expected epithermal neutron population escaping our $^3$He proportional counter sensitivity operated three meters away from the IP2I cryostat. 
Indeed, it is worth noting that in such configuration the bare $^3$He is probing almost exclusively the  neutron evaporation peak (see  Fig.~\ref{fig:He3Fluxes}), while the bolometers are also sensitive to the cascade peak as the neutrons get down-converted to lower energies thanks to the lead shielding surrounding the cryostat. 
Such neutrons would then induce a larger than expected keV-scale nuclear recoil rate in our bolometers. We plan to test this hypothesis using lithiated bolometers~\cite{Coron:2012eq,Armengaud:2017hit}, operated in our cryostat at IP2I, and hydrogen recoil proportional counters which should exhibit complementary epithermal neutron sensitivity to our $^3$He detector.
%This hypothesis will be further tested with the use of lithiated bolometers for an in-situ neutron background monitoring of the future \Ricochet{} experiment~\cite{billard:tel-03259707}.

  \begin{figure*}
\begin{center}
\includegraphics[width=\columnwidth,angle=0]{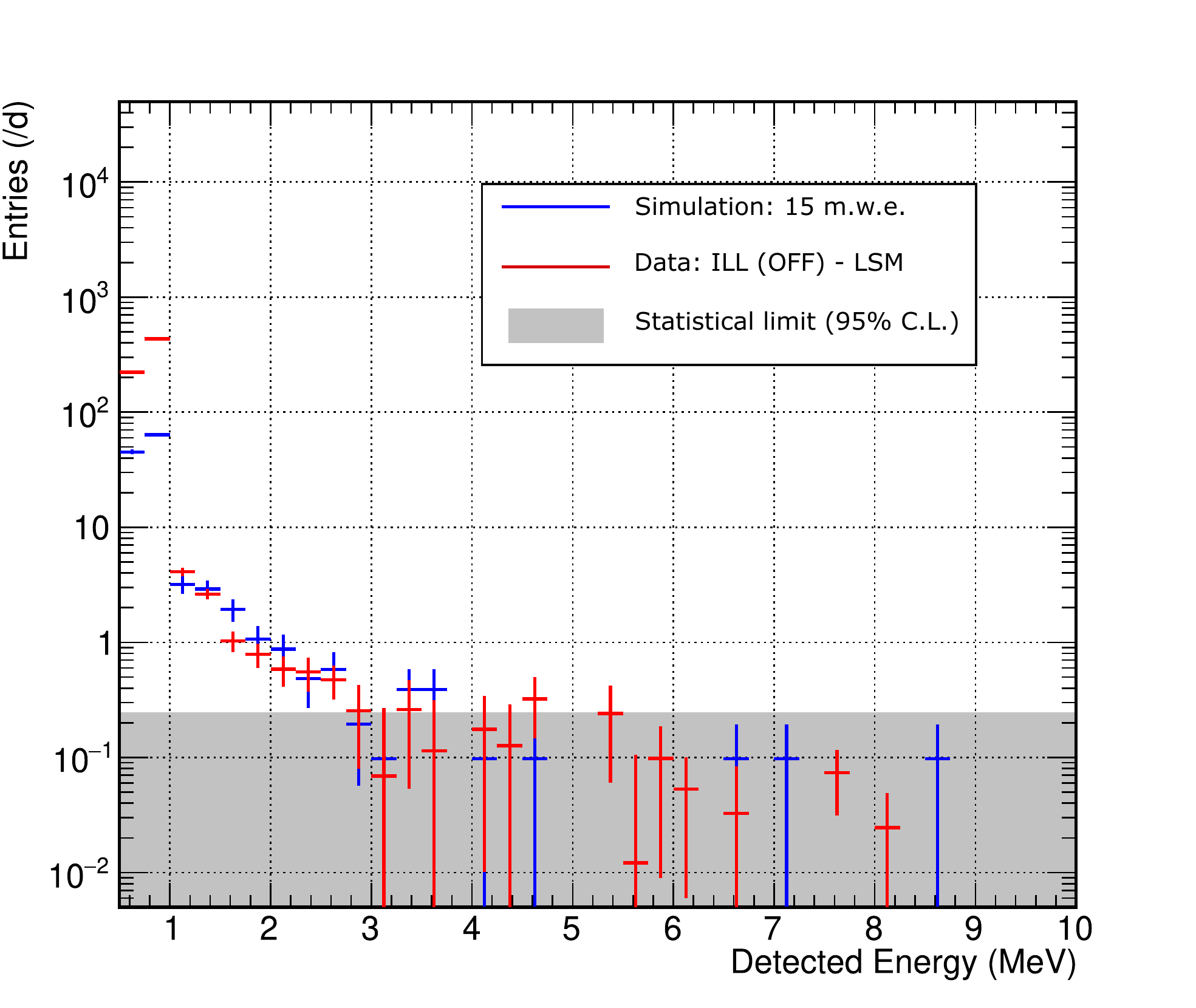}
\includegraphics[width=\columnwidth,angle=0]{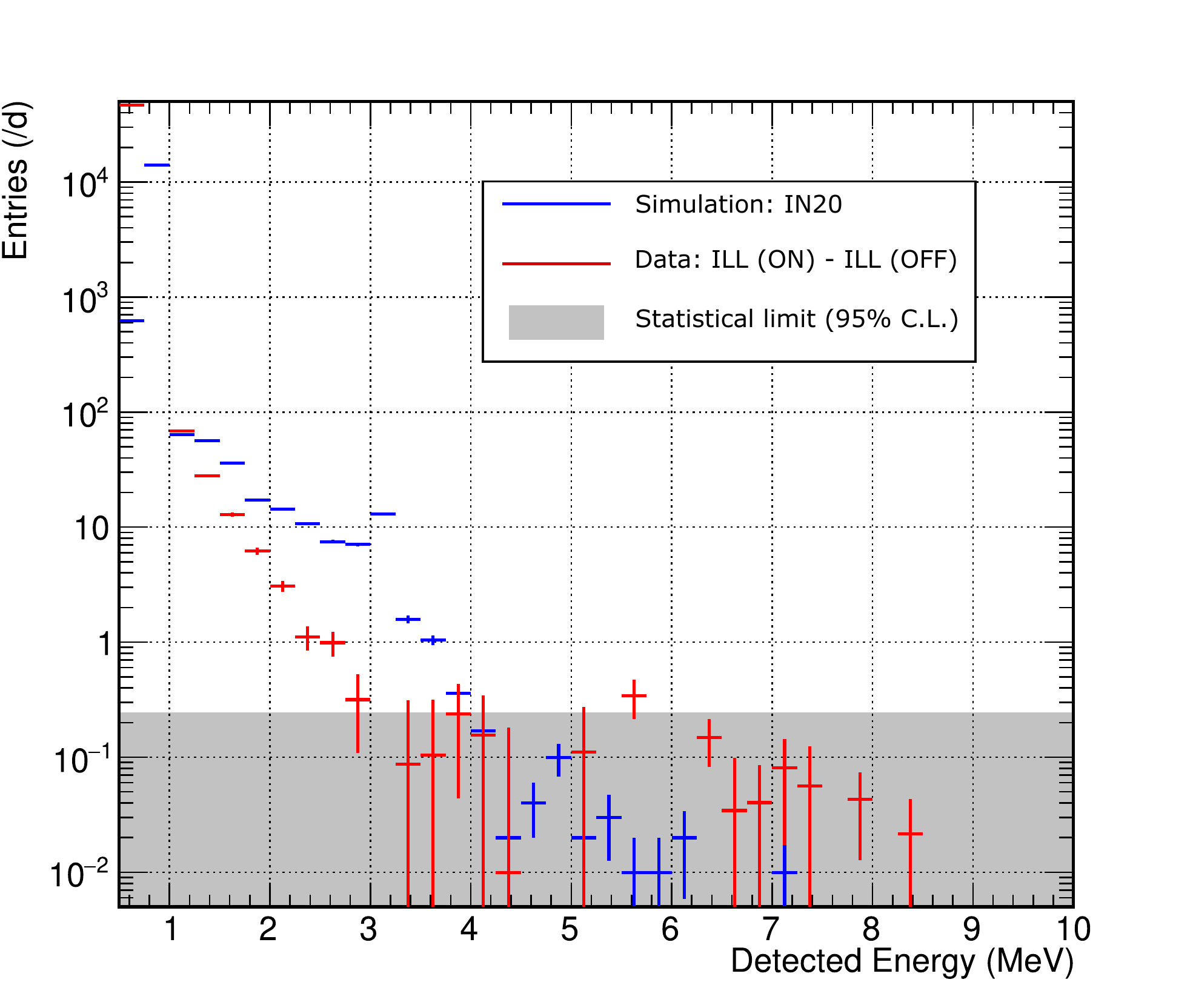}
\caption{{\bf Left:} Comparison between the observed $^3$He energy spectra of the cosmogenic neutron background at the ILL-H7 site after 40.7 days of data taking (red) and its corresponding simulations (blue). Note that the red histogram has been obtained by subtracting the LSM data in order to remove the internal alpha background contamination. {\bf Right:} Comparison between the observed $^3$He energy spectra of the reactogenic neutron background at the ILL-H7 site after 46.3 days of data taking with an averaged reactor power of 42~MW (red) and its corresponding simulations (blue). Note that the red histogram has been obtained by subtracting the reactor OFF data, which subtracts both the residual internal background contamination and the cosmogenic component within statistical uncertainties. All measurements at the ILL reactor were done with a 1~cm thick layer of boron-loaded rubber around the $^3$He proportional counter. Lastly, as discussed in Sec.~\ref{sec:He3counter}, due to the internal proportional counter background, a statistically significant neutron contribution can be inferred from an observed rate larger than about 0.25~evt/day/bin at 95\% C.L.} 
\label{fig:ResultsILL}
\vspace{-0.5cm}
\end{center}
\end{figure*}

Eventually, the qualitative concordance between the simulated and observed nuclear recoil background with the Ge bolometers confirms the reliability of our proposed neutron background assessment approach using a $^3$He proportional counter combined with both muon flux attenuation measurements and CRY-based simulations. Quantitatively it appears that in the case of our cryogenic lab at IP2I, with a measured $\sim$3~m.w.e. overburden and no polyethylene shielding around the cryostat, we are underestimating the neutron background at the Ge bolometers by a factor of about three. For the sake of completness, we will therefore consider this re-scaling factor as worst case scenario of our \Ricochet{} sensitivity study presented in Sec.~\ref{sec:RicochetBackground}.

%Even though this source of discrepancy is still under investigation, it is worth noticing that epithermal neutrons will be well shielded against in the future \Ricochet{} experiment thanks to its 35~cm thick polyethylene shielding layer. 
%  In the future, we plan to improve our proposed neutron background assessment approach in expending it with additional neutron detectors such as plastic scintillators~\cite{Angloher:2019flc} and lithiated bolometers~\cite{} to improve our onsite neutron background monitoring for the future \Ricochet experiment at ILL.

\subsection{\Ricochet{} fast neutron background characterization: cosmogenic and reactogenic neutrons at the ILL-H7 site}
\label{sec:ILLneutronMeasure}

At the end of 2020, the STEREO experiment was decommissioned. Since then, the ILL-H7 site has been empty and therefore perfectly well-suited for background and on-site characterizations prior to the \Ricochet{} integration. Starting in January 2021, we took almost a hundred days worth of data, during reactor ON and OFF periods, with the $^3$He proportional counter located at the planned position of the \Ricochet{} cryostat. To properly simulate the ILL site, we used the altitude and latitude of Grenoble which are 212~m above sea level and 45$^\circ$11’18.704” North, respectively, and also applied the 1.28 cosmic flux normalization factor from STEREO (see  Sec.~\ref{sec:simuneutronflux}).

Figure~\ref{fig:ResultsILL} (left panel) shows the resulting comparison between the cosmogenic simulations (blue) and the observed data (red) of the $^3$He detector at the ILL-H7 site when the reactor is OFF. Similarly to the IP2I case, the red histogram has been obtained after subtraction of the event rate observed from the LSM data in order to subtract the residual internal background. Again, an excellent agreement between the experimental data and the cosmogenic simulations is observed above 1~MeV in detected energy, hence validating our cosmogenic neutron flux model to be used to estimate the corresponding neutron background to the future \Ricochet{} experiment.

Data with our proportional counter was also acquired during reactor ON periods in order to estimate the reactogenic neutron flux. Figure~\ref{fig:ResultsILL} (right panel) presents the resulting  reactogenic neutron data and simulations. The experimental data (red histogram) has been derived by subtracting the OFF period to remove both the cosmogenic neutrons and the residual internal background contributions. The simulated spectrum (blue) has been obtained by scaling the spectrum from IN20 in Fig.~\ref{fig:He3Fluxes} (red histogram) to the reduced 42~MW thermal power during our measurements.  First, it is worth noticing that we observe a fast neutron detection rate about 10 times higher during reactor ON periods ($121.9 \pm 1.9$ per day) with respect to OFF periods ($11.5 \pm 0.9$ per day), for that reactor power of 42~MW and IN20 in operation. Taken at face value, this suggests an overall reactogenic fast neutron flux about 15 times higher than the cosmogenic one when the reactor is operated at its full 58~MW nominal thermal power. Note that a higher reactogenic fast neutron flux is also expected from Fig.~\ref{fig:He3Fluxes}. Also, in this case we observe a significant departure between the two histograms, suggesting that our simulated reactogenic neutron flux is both too high and at higher energies than what we observe. Similarly, our simulations predict a fast neutron detection rate of about 230 per day above 1~MeV, hence almost two times higher than the observed one. This difference can be explained by the fact that the IN20 neutron spectrum considered here doesn't take into account the lead and polyethylene walls that are surrounding the ILL-H7 site, nor the neutron moderator and shielding from the IN20 instrument. As suggested in Sec.~\ref{sec:simuneutronflux}, it was indeed expected that our neutron background model assumption, using the outgoing IN20 reactogenic neutron flux from the H13 beam, would overestimate the fast neutron flux at the \Ricochet{} location. However, in order to provide some conservative estimates of the expected neutron background, we consider hereafter this un-moderated IN20 neutron flux as an input to our \Ricochet{} background simulations.

\section{\Ricochet{} expected neutron background}
\label{sec:RicochetBackground}

%%%%%% Table %%%%%%%%%

\newcommand{\TitleTabER}{\multirow{3}{*}{\begin{tabular}[c]{@{}c@{}}Electronic recoils \\ {[}50\,eV, 1\,keV{]} \\ (evts/day/kg)\end{tabular}}}
\newcommand{\TitleTabNR}{\multirow{3}{*}{\begin{tabular}[c]{@{}c@{}}Nuclear recoils    \\ {[}50\,eV, 1\,keV{]} \\ (evts/day/kg)\end{tabular}}}

\newcommand{\CosmoNoShieldER}{$260 \pm 5$}                   \newcommand{\ReactoNoShieldER}{$4365 \pm 301$}               \newcommand{\SumNoShieldER}{$4625 \pm 301$}                 
\newcommand{\CosmoShieldER}{$183 \pm 6$}                     \newcommand{\ReactoShieldER}{$18\pm 2$}                     \newcommand{\SumShieldER}{$201 \pm 6$}                      
\newcommand{\CosmoShieldVetoER}{$1.6\pm0.6$}                                                                            \newcommand{\SumShieldVetoER}{$20\pm2$}                   

\newcommand{\CosmoNoShieldNR}{$1554 \pm 12$}                 \newcommand{\ReactoNoShieldNR}{$53853 \pm 544$}              \newcommand{\SumNoShieldNR}{$55407 \pm 545$}                
\newcommand{\CosmoShieldNR}{$42\pm 3$}                      \newcommand{\ReactoShieldNR}{$2.4\pm0.3$}                  \newcommand{\SumShieldNR}{$44\pm3$}                       
\newcommand{\CosmoShieldVetoNR}{$7\pm2$}                                                                               \newcommand{\SumShieldVetoNR}{$9\pm2$}

\begin{table*}[t]
\centering
\smallskip
\def\arraystretch{1.5}
\begin{tabular}{@{}cccccc@{}}
\toprule[1pt]
\multicolumn{2}{c}{}                            & Cosmogenic         & Reactogenic                       & Total (MC)       & {\bf CENNS (Ge/Zn)}   \\
\midrule[0.5pt]
 \TitleTabNR & No Shielding  (I)                   & \CosmoNoShieldNR   & \ReactoNoShieldNR                 & \SumNoShieldNR   & --                          \\
             & Passive Shielding (II)               & \CosmoShieldNR     & \multirow{2}{*}{\ReactoShieldNR}  & \SumShieldNR     & --                           \\
             & Passive + $\mu$-veto (III) & \CosmoShieldVetoNR &                                   & \SumShieldVetoNR & {\bf 12.8 / 11.2}             \\
\bottomrule[1pt]
\end{tabular}
\caption{\label{tab:SimuRate}Simulated background rates inside the cryogenic detector array installed at the ILL, with the shielding design illustrated in Fig.~\ref{fig:RicochetSetup}, when only one bolometer has triggered. As the muon veto is still being characterized and optimized, in the case of scenario (III) we assume perfect geometrical and detection efficiencies.}
\end{table*}

%%%%%%% End Table %%%%%%

From the cosmogenic and reactogenic neutron components of the expected \Ricochet{} background -- compared against the $^3$He counter data in the previous section (see Sec.~\ref{sec:NeutronFluxMeasure}) --  we can estimate the expected \Ricochet{} neutron background using a GEANT4 simulation taking into account its entire shielding and detector geometry, introduced in Sec.~\ref{sec:Ricochet}.

Table~\ref{tab:SimuRate} presents the resulting expected reactogenic and cosmogenic neutron background rates, integrated over our CENNS region of interest between 50~eV and 1~keV, for various shielding configurations: (I) no shielding, (II) with the passive shielding presented in Fig.~\ref{fig:RicochetSetup}, and (III) with the addition of an idealized muon veto assumed to have a 100\% geometrical and detection efficiency surrounding the \Ricochet{} experimental setup. From the comparison of the first two shielding configurations I and II, one can derive that the neutron background attenuation factors provided by the passive \Ricochet{} shielding are about 37 and of the order of $10^4$ for the cosmogenic and reactogenic neutron backgrounds, respectively. 
%The much greater attenuation factor for reactogenic neutrons is easily explained by their comparatively lower energy.
%The much greater attenuation factor for reactogenic neutrons is explained by their comparatively low energy when compared to that of both primary and muon-induced neutrons from the cosmogenic contribution.
The much greater attenuation factor for reactogenic neutrons is  explained by both 1) the absence of muon-induced spallation in the shielding producing fast neutrons in close proximity to the detectors, and 2) their comparatively low energy when compared to that of primary and spallation neutrons from the cosmogenic contribution as they enter the \Ricochet{} shielding.
Indeed, most of these reactogenic neutrons have kinetic energies below 6~MeV (see Fig.~\ref{fig:He3Fluxes}), corresponding to a mean free path in polyethylene of about 6~cm, making them efficiently moderated by the 35~cm of polyethylene. On the other hand, with energies up to $\sim$200~MeV, cosmogenic neutrons can still reach the \Ricochet{} cryogenic detectors. Therefore, despite of their higher expected (and measured) overall fast neutron flux, reactogenic neutrons are not expected to be a dominant background to the future \Ricochet{} experiment even when considering the extreme case of the un-moderated IN20 simulated neutron flux (see Sec.~\ref{sec:ILLneutronMeasure}). However, note that reaching such high attenuation factors puts strong constraints on the tightness of the passive shielding, hence the additional internal layers between the thermal screens to limit possible neutron leakage to the bolometers from the top (see Sec.~\ref{sec:Ricochet}).

The comparison of the shielding configurations II and III from Table~\ref{tab:SimuRate} suggests that an idealized muon veto could help reducing the cosmogenic neutron background by an additional factor of 6. As the \Ricochet{} muon veto won't be as efficient as an ideal one, we indeed expect an overall muon veto tagging efficiency of about 90\%, we consider hereafter that our cosmogenic neutron background will be between $42\pm3$ and $7\pm2$ events per day.  Solely considering the expected neutron backgrounds, these two cases respectively lead to signal-to-background ratios of about 0.3 and 1.4. Assuming a 70\% CENNS detection efficiency, arising from estimated livetime loss and of various analysis cuts finite efficiencies, these values suggest that the \Ricochet{} experiment could reach a statistical CENNS detection significance\footnote{The significance is defined as $Z = S/\sqrt{(S+2B)}$ with $S$ and $B$ the numbers of CENNS and background events respectively and assuming equal reactor ON and OFF exposition times.} after only one reactor cycle between 7.5~$\sigma$ and  13.6~$\sigma$. If we apply a conservative factor of 3 to the neutron background rates based on the Ge bolometer comparison between our cosmogenic simulations and observations done at IP2I (see Sec.~\ref{sec:ValidationIP2I}), these significances drop to 4.6~$\sigma$ and 9.2~$\sigma$ depending on the muon veto efficiency, respectively. Lastly, it is worth highlighting that these neutron background based sensitivity estimates assume that there are no additional unexpected backgrounds, and that the gamma background will be both low enough and efficiently rejected thanks to the \Ricochet{} bolometers' particle identification capabilities.

\section{Conclusion}

In this paper, we have presented our fast neutron flux characterization with a dedicated low-background $^3$He proportional counter. We first tested our method in comparing simulated and observed energy spectra from the IP2I surface lab where cryogenic detectors, with particle identification capabilities, were also operated. This allowed us to cross-check that our cosmogenic simulations were properly reproducing both the $^3$He spectra above 1~MeV in detected energy and the low-energy nuclear recoil spectrum from our Ge bolometers to within a factor of about three, assuming a sole cosmogenic neutron component. Following this cross-validation, we measured and simulated the neutron fluxes at the ILL-H7 site, where the future CENNS \Ricochet{} experiment will be deployed. Firstly, we found an excellent agreement between our cosmogenic neutron simulation and measurements. Based on these observations, one can conclude that CRY provides reliable estimates of cosmogenic backgrounds for experiments located at shallow sites with depths from 3 to 15~m.w.e. Secondly, a significant  disagreement has been found in the case of reactogenic neutrons between our $^3$He simulations and experimental data, suggesting that the IN20 neutron model considered here overestimates the reactor induced neutron energies and flux at the ILL-H7 site. Therefore, the IN20 neutron flux model is considered as a conservative model to estimate the anticipated reactogenic neutron background for \Ricochet{}. Following these onsite neutron background characterizations, we propagated both our reactogenic and cosmogenic neutron fluxes into our \Ricochet{} shielding simulation to estimate its expected nuclear recoil background level. Interestingly, despite its higher fast neutron flux, we found that the reactogenic neutron background will only contribute to about one forth of the overall \Ricochet{} neutron background, suggesting that the ON/OFF reactor modulations should lead to an increased CENNS sensitivity. Assuming our neutron background model, we found that depending on the effectiveness of the muon veto, a statistical significance of a CENNS detection with \Ricochet{} after only one reactor cycle should be between 7.5 to 13.6~$\sigma$ when only the expected neutron background is considered. A similar study dedicated to the gamma induced background, also addressing the particle identification capabilities of our detectors, is ongoing and will be presented in a forthcoming paper.

\section*{Acknowledgments}

We are grateful to the EDELWEISS collaboration for the use of its electronics and DAQ system in the operation of the RED20 and RED80 cryogenic detectors discussed in Sec.~\ref{sec:NeutronFluxMeasure}. This project has received funding from the European Research Council (ERC) under the European Union’s Horizon 2020 research and innovation program under Grant Agreement ERC-StG-CENNS 803079, the French National Research Agency (ANR) within the project ANR-20-CE31-0006, the LabEx Lyon Institute of Origins (ANR-10-LABX-0066) of the Université de Lyon, the NSF under Grant PHY-2013203.  A portion of the work carried out at MIT was supported by DOE QuantISED award DE-SC0020181 and the Heising-Simons Foundation. This work is also partly supported by the Ministry of science and higher education of the Russian Federation (the contract No. 075-15-2020-778).

\bibliographystyle{spphys} 
\bibliography{Refs}

\end{document}